# New scenario for transition to slow 3-D turbulence

# Part I. Slow 1-D turbulence in Nikolaevskii system.

J. Foukzon

Israel Institute of Technology, Haifa, Israel

jaykovfoukzon@list.ru

***Abstract:*** Analytical non-perturbative study of the three-dimensional nonlinear stochastic partial differential equation with additive thermal noise, analogous to that proposed by V.N. Nikolaevskii [1]-[5]to describe longitudinal seismic waves, is presented. The equation has a threshold of short-wave instability and symmetry, providing long wave dynamics. New mechanism of quantum chaos generating in nonlinear dynamical systems with infinite number of degrees of freedom is proposed. The hypothesis is said, that physical turbulence could be identified with quantum chaos of considered type. It is shown that the additive thermal noise destabilizes dramatically the ground state of the Nikolaevskii system thus causing it to make a direct transition from a spatially uniform to a turbulent state.

**1.Introduction.** In the present work a non-perturbative analytical approach to the studying of problem of quantum chaos in dynamical systems with infinite number of degrees of freedom is proposed. Statistical descriptions of dynamical chaos and investigations of noise effects on chaotic regimes are studied. Proposed approach also allows estimate the influence of additive (thermal)fluctuations on the processes of formation of developed turbulence modes in essentially nonlinear processes like electro-convection and other. A principal role the influence of thermal fluctuations on the dynamics of some types of dissipative systems in the approximate environs of derivation rapid of a short-wave instability was ascertained. Impotent physical results follows from Theorem 2, is illustrated by example of 3D stochastic model system:

$$\frac{\partial u_\eta(x,t,\varepsilon)}{\partial t}+\Delta[\varepsilon-(1+\Delta)^2]u_\eta(x,t,\varepsilon)+\left[\delta_1\frac{\partial u_\eta(x,t,\varepsilon)}{\partial x_1}+\delta_2\frac{\partial u_\eta(x,t,\varepsilon)}{\partial x_2}+\delta_3\frac{\partial u_\eta(x,t,\varepsilon)}{\partial x_3}\right]u_\eta(x,t,\varepsilon)+$$

$$+f(x,t)-\sqrt{\eta}w(x,t)=0,\quad x\in\mathbb{R}^3 \qquad (1.1)$$

$$u_\eta(x,0,\varepsilon)=0, w(x,t)=\frac{\partial^4 W(x,t)}{\partial x_1\partial x_2\partial x_3\partial t},\eta\ll 1,0<\delta_j,j=1,2,3, \qquad (1.2)$$

which was obtained from the non-stochastic $3D$ Nikolaevskii model:

$$\frac{\partial u(x,t,\varepsilon)}{\partial t} + \Delta[\varepsilon - (1+\Delta)^2]u(x,t,\varepsilon) + \left[\delta_1 \frac{\partial u(x,t,\varepsilon)}{\partial x_1} + \delta_2 \frac{\partial u(x,t,\varepsilon)}{\partial x_2} + \delta_3 \frac{\partial u(x,t,\varepsilon)}{\partial x_3}\right] u(x,t,\varepsilon) + f(x,t) \quad (1.3)$$

which is perturbed by additive "small" white noise $\sqrt{\eta}w(x,t)$. And analytical result also illustrated by example of$1D$ stochastic model system

$$\frac{\partial u_\eta(x,t,\varepsilon)}{\partial t} + \Delta[\varepsilon - (1+\Delta)^2]u_\eta(x,t,\varepsilon) + \delta \frac{\partial u_\eta(x,t,\varepsilon)}{\partial x_1} u_\eta(x,t,\varepsilon) + f(x,t) - \sqrt{\eta}w(x,t) = 0, x \in \mathbb{R} \quad (1.4)$$

$$u_\eta(x,0,\varepsilon) = 0, w(x,t) = \frac{\partial^2 W(x,t)}{\partial x \partial t}, \eta \ll 1, 0 < \delta, \quad (1.5)$$

which was obtained from the non-stochastic 1D Nikolaevskii s model:

$$\frac{\partial u(x,t,\varepsilon)}{\partial t} + \Delta[\varepsilon - (1+\Delta)^2]u(x,t,\varepsilon) + \delta \frac{\partial u(x,t,\varepsilon)}{\partial x_1} u(x,t,\varepsilon) + f(x,t) = 0, u(x,0,\varepsilon) = 0, x \in \mathbb{R}, \quad (1.6)$$

$$u(x,0,\varepsilon) = 0. \quad (1.7)$$

Systematic study of a different type of chaos at onset ''soft-mode turbulence'' based on numerical integration of the simplest 1D Nikolaevskii model (1.7) has been executed by many authors [2]-[7].There is an erroneous belief that such numerical integration gives a powerful analysis is means of the processes of turbulence conception, based on the classical theory of chaos of the finite-dimensional classical systems [8]-[11].

**Remark1.1.** However, as it well known, such approximations correct only in a phase of turbulence conception, when relatively small number of the degrees of freedom excites**.** In general case, when a very large number of the degrees of freedom excites, well known phenomena of the numerically induced ***c***haos, can to spoils in the uncontrollable way any numerical integration [12]-[15]

**Remark1.2.** Other non trivial problem stays from noise roundoff error in computer computation using floating point arithmetic [16]-[20]. In any computer simulation the numerical solution is fraught with truncation by roundoff errors introduced by finite-precision calculation of trajectories of dynamical systems, where roundoff errors or other noise can introduce new behavior and this problem is a very more pronounced in the case of chaotic dynamical systems, because the trajectories of such systems exhibit extensive dependence on initial conditions. As a result, a small random truncation or roundoff error, made computational error at any step of computation will tend to be large magnified by future computational of the system [17].

**Remark1.3.** As it well known, if the digitized or rounded quantity is allowed to occupy the nearest of a large number of levels whose smallest separation is $E_0$, then, provided that the original quantity is large compared to $E_0$ and is reasonably well behaved, the effect of the quantization or rounding may be treated as additive random noise [18].Bennett has shown that such additive noise is nearly white, with mean squared value of $E_0^2/12$ [19].However the complete uniform white-noise model to be valid in the sense of weak convergence of probabilistic measures as the lattice step tends to zero if the matrices of realization of the system in the state space satisfy certain non resonance conditions and the finite-dimensional distributions of the input signal are absolutely continuous [19].

The method deprived of these essential lacks in general case has been offered by the author in papers [23]-[27].

**Remark1.4 .** Thus from consideration above it is clear that numerical integration procedure of the1D Nikolaevskii model (1.6)-(1.7) executed in papers [2]-[7] in fact dealing with stochastic model (1.4)-(1.5). There is an erroneous the point of view, that a white noise with enough small intensity does not bring any significant contributions in turbulent modes, see for example [3]. By this wrong assumptions the results of the numerical integration procedure of the1D Nikolaevskii model (1.6)-(1.7) were mistakenly considered and interpreted as a very exact modeling the slow turbulence within purely non stochastic Nikolaevskii model (1.6)-(1.7). Accordingly wrong conclusions about that temperature noises does not influence slow turbulence have been proposed in [3].However in [27] has shown non-perturbatively that that a white noise with enough small intensity can to bring significant contributions in turbulent modes and even to change this modes dramatically.

At the present time it is generally recognized that turbulence in its developed phase has essentially singular spatially-temporal structure. Such a singular conduct is impossible to describe adequately by the means of some model system of equations of a finite dimensionality. In this point a classical theory of chaos is able to describe only small part of turbulence phenomenon in liquid and another analogous s of dynamical systems. The results of non-perturbative modeling of super-chaotic modes, obtained in the present paper allow us to put out a quite probable hypothesis: developed turbulence in the real physical systems with infinite number of degrees of freedom is a quantum super-chaos, at that the quantitative characteristics of this super-chaos, is completely determined by non-perturbative contribution of additive (thermal) fluctuations in the corresponding classical system dynamics [18]-[20].

## 2.Main Theoretical Results

We study the stochastic $r$-dimensional differential equation analogous proposed by Nikolaevskii [1] to describe longitudinal seismic waves:

$$\frac{\partial u_\eta(x,t,\varepsilon,\omega)}{\partial t}+\Delta[\varepsilon-(1+\Delta)^2]u_\eta(x,t,\varepsilon,\omega)+u_\eta(x,t,\varepsilon,\omega)\sum_{j=1}^{r}\delta_i\frac{\partial u_\eta(x,t,\varepsilon,\omega)}{\partial x_i}+f(x,t)-\sqrt{\eta}w(x,t,\omega)=0,\ (2.1)$$

$$x\in\mathbb{R}^r, u_\eta(x,0,\varepsilon,\omega)=0, w(x,t)=\frac{\partial^{r+1}W(x,t)}{\partial x_1\partial x_2\ldots\partial x_r\partial t}, 0<\delta_j, j=1,\ldots,r.(2.2)$$

The main difficulty with the stochastic Nikolaevskii equation is that the solutions do not take values in an function space but in generalized function space. Thus it is necessary to give meaning to the non-linear terms $\partial_{x_j}u_\eta^2(x,t,\varepsilon,\omega), j=1,\ldots,r$ because the usual product makes no sense for arbitrary distributions. We deal with product of distributions via regularizations, i.e., we approximate the distributions by appropriate way and pass to the limit. In this paper we use the approximation of the distributions by approach of Colombeau generalized functions [28].

**Notation 2.1.** We denote by $\mathcal{D}(\mathbb{R}^r\times\mathbb{R}_+)$ the space of the infinitely differentiable functions with compact support in $\mathbb{R}^r\times\mathbb{R}_+$ and by $\mathcal{D}'(\mathbb{R}^r\times\mathbb{R}_+)$ its dual space. Let $\mathfrak{C}=(\Omega,\Sigma,\mu)$ be a probability space. We denote by $\mathbf{D}$ the space of all functions $T:\Omega\to\mathcal{D}'(\mathbb{R}^r\times\mathbb{R}_+)$ such that $\langle T,\varphi\rangle$ is a random variable for all $\varphi\in\mathcal{D}(\mathbb{R}^r\times\mathbb{R}_+)$. The elements of $\mathbf{D}$ are called random generalized functions.

**Definition 2.1.** [29]. We say that a random field $\{\mathfrak{R}(x,t)|t\in\mathbb{R}_+, x\in\mathbb{R}^r\}$ is a spatially dependent semimartingale if for each $x\in\mathbb{R}^r, \{\mathfrak{R}(x,t)|t\in\mathbb{R}_+\}$ is a semimartingale in relation to the same filtration $\{\mathcal{F}_t|t\in\mathbb{R}_+\}$. If $\mathfrak{R}(x,t)$ is a $C^\infty$-function of $x$ and continuous in almost everywhere, it is called a $C^\infty$-semimartingale.

**Definition 2.2.** We say that that $u_\eta(x,t,\varepsilon,\omega)\in\mathbf{D}$ is a strong generalized solution (**SGS**) of the Eq.(2.1)-(2.2) if there exists a sequence of $C^\infty$-semimartingales $u_\eta(x,t,\varepsilon,\epsilon,\omega), \epsilon\in(0,1]$ such that there exists

(i) $u_\eta(x,t,\varepsilon,\omega)=_{\text{def}}\lim_{\epsilon\to0}u_\eta(x,t,\varepsilon,\epsilon,\omega)$ in $\mathcal{D}'(\mathbb{R}^r\times\mathbb{R}_+)$ almost surely for $\omega\in\Omega$,

(ii) $\partial_{x_j}u_\eta^2(x,t,\varepsilon,\epsilon,\omega)=_{\text{def}}\lim_{\epsilon\to0}\partial_{x_j}u_\eta^2(x,t,\varepsilon,\epsilon,\omega), j=1,\ldots,r$ almost surely for $\omega\in\Omega$,

(iii) for all $\varphi\in\mathcal{D}(\mathbb{R}^r\times\mathbb{R}_+)$

(iv) $\langle\partial_t u_\eta(x,t,\varepsilon,\omega),\varphi\rangle-\langle\Delta[\varepsilon-(1+\Delta)^2]u_\eta(x,t,\varepsilon,\omega),\varphi\rangle-\sum_{j=1}^{r}\frac{\delta_i}{2}\langle\partial_{x_j}u_\eta^2(x,t,\varepsilon,\omega),\varphi\rangle-\langle f(x,t),\varphi\rangle+$

$+\sqrt{\eta}\int_{\mathbb{R}^r} dx \int_0^\infty \varphi(t,x)\, dW_t(x,t) = 0,\ \ t \in \mathbb{R}_+$ almost surely for $\omega \in \Omega$,

and where $W_t(x,t) = \frac{\partial^r W(x,t)}{\partial x_1 \partial x_2 \ldots \partial x_r}$,

(v) $u_\eta(x,t,\varepsilon,\omega) = 0$ almost surely for $\omega \in \Omega$.

However in this paper we use the solutions of stochastic Nikolaevskii equation only in the sense of Colombeau generalized functions [30].

**Remark2.1.** Note that from Definition 2.2 it is clear that any strong generalized solution $u_\eta(x,t,\varepsilon,\omega)$ of the Eq.(2.1)-(2.2) one can to recognized as Colombeau generalized function such that

$$u_\eta(x,t,\varepsilon,\omega) =_{\text{def}} \left(u_\eta(x,t,\varepsilon,\epsilon,\omega)\right)_\epsilon \quad (\#)$$

By formula (#) one can to define appropriate generalized solution of the Eq.(2.1)-(2.2) even if a strong generalized solution of the Eq.(2.1)-(2.2) does not exist.

**Definition 2.3.** Assume that a strong generalized solution of the Eq.(2.1)-(2.2) does not exist. We shall say that:

(I) Colombeau generalized stochastic process $\left(u_\eta(x,t,\varepsilon,\epsilon,\omega)\right)_\epsilon$ is a weak generalized solution (**WGS**) of the Eq.(2.1)-(2.2) or Colombeau solution of the Eq.(2.1)-(2.2) if for all $\varphi \in \mathcal{D}(\mathbb{R}^r \times \mathbb{R}_+)$ and for all $\epsilon \in (0,1]$

(i) $$\langle u_\eta(x,t,\varepsilon,\epsilon,\omega), \partial_t\varphi\rangle - \langle \Delta[\varepsilon - (1+\Delta)^2]u_\eta(x,t,\varepsilon,\epsilon,\omega), \varphi\rangle - \sum_{j=1}^r \frac{\delta_i}{2}\langle \partial_{x_j} u_\eta^2(x,t,\varepsilon,\epsilon,\omega), \varphi\rangle +$$
$$+ \langle f(x,t), \varphi\rangle + \sqrt{\eta}\int_{\mathbb{R}^r} dx \int_0^\infty \varphi(t,x)\, dW_t(x,t) = 0,\ \ t \in \mathbb{R}_+$$ almost surely for $\omega \in \Omega$,

(ii) $u_\eta(x,t,\varepsilon,\epsilon,\omega) = 0$ almost surely for $\omega \in \Omega$.

(**II**) Colombeau generalized stochastic process $\left(u_\eta(x,t,\varepsilon,\epsilon,\omega)\right)_\epsilon$ is a Colombeau-Ito's solution of the Eq.(2.1)-(2.2) if for all $\varphi \in \mathcal{D}(\mathbb{R}^r)$ and for all $\epsilon \in (0,1]$

(i) $$\langle \partial_t u_\eta(x,t,\varepsilon,\epsilon,\omega), \varphi\rangle + \langle \Delta[\varepsilon - (1+\Delta)^2]u_\eta(x,t,\varepsilon,\epsilon,\omega), \varphi\rangle \sum_{j=1}^r \frac{\delta_i}{2}\langle \partial_{x_j} u_\eta^2(x,t,\varepsilon,\epsilon,\omega), \varphi\rangle -$$
$$-\langle f(x,t), \varphi\rangle - \sqrt{\eta}\int_{\mathbb{R}^r} \varphi(x)\, w(x,t) dx = 0,\ \ t \in \mathbb{R}_+$$ almost surely for $\omega \in \Omega$,

(ii) $u_\eta(x,t,\varepsilon,\epsilon,\omega) = 0$ almost surely for $\omega \in \Omega$.

**Notation 2.2.** [30]. The algebra of moderate element we denote by $\mathcal{E}_M[\mathbb{R}^r]$. The Colombeau algebra of the Colombeau generalized function we denote by $\mathcal{G}(\mathbb{R}^r)$.

**Notation 2.3.** [30].We shall use the following designations. If $U \in \mathcal{G}(\mathbb{R}^r)$ it representatives will be denoted by $R_U$, their values on $\varphi = (\varphi_\epsilon(x)), \epsilon \in (0,1]$ will be denoted by $R_U(\varphi)$ and it point values at $x \in \mathbb{R}^r$ will be denoted $R_U(\varphi, x)$.

**Definition 2.4.** [30]. Let $A_0 = A_0(\mathbb{R}^r)$ be the set of all $\varphi \in D(\mathbb{R}^r)$ such that $\int \varphi(x)\,dx = 1$.
Let $\mathfrak{C} = (\Omega, \Sigma, \mu)$ be a probability space. Colombeau random generalized function this is a map $U: \Omega \to \mathcal{G}(\mathbb{R}^r)$ such that there is representing function $R_U$: $A_0 \times \mathbb{R}^r \times \Omega$ with the properties:

(i) for fixed $\varphi \in A_0(\mathbb{R}^r)$ the function $(x, \omega) \to R_U(\varphi, x, \omega)$ is a jointly measurable on $\mathbb{R}^r \times \Omega$;
(ii) almost surely in $\omega \in \Omega$, the function $\varphi \to R_U(\varphi, ., \omega)$ belongs to $\mathcal{E}_M[\mathbb{R}^r]$ and is a representative of $U$;

**Notation 2.3.** [30]. The Colombeau algebra of Colombeau random generalized function is denoted by $\mathcal{G}_\Omega(\mathbb{R}^r)$.

**Definition 2.5.** Let $\mathfrak{C} = (\Omega, \Sigma, \mu)$ be a probability space. Classically, a generalized stochastic process on $\mathbb{R}^r$ is a weakly measurable map $V: \Omega \to D'(\mathbb{R}^r)$ denoted by $V \in D'_\Omega(\mathbb{R}^r)$. If $\varphi \in A_0(\mathbb{R}^r)$, then

(iii) $V(\omega) * \varphi(x) = \langle V(\omega), \varphi(x-.)\rangle$ is a measurable with respect to $\omega \in \Omega$ and
(iv) smooth with respect to $x \in \mathbb{R}^r$ and hence jointly measurable.
(v) Also $\left(V(\omega) * \varphi(x)\right) \in \mathcal{E}_M[\mathbb{R}^r]$.
(vi) Therefore $R_V(\varphi, x, \omega) = V(\omega) * \varphi(x)$ qualifies as an representing function for an element of $\mathcal{G}_\Omega(\mathbb{R}^r)$.
(vii) In this way we have an imbedding $\mathcal{D}'(\mathbb{R}^r) \to \mathcal{G}_\Omega(\mathbb{R}^r)$.

**Definition 2.6.** Denote by $S(T) = S(\mathbb{R}^{r+1}) \upharpoonright T$ the space of rapidly decreasing smooth functions on $T = \mathbb{R}^r \times [0, \infty)$. Let $\mathfrak{C} = (\Omega, \Sigma, \mu)$ with (i) $\Omega = S'(T)$, (ii) $\Sigma$- the Borel $\sigma$-algebra generated by the weak topology. Therefore there is unique probability measure $\mu$ on $(\Omega, \Sigma)$ such that

$$\int d\mu(\omega) \exp\left[i\langle\omega, \varphi\rangle\right] = \exp\left(-\frac{1}{2}\|\varphi\|^2_{L_2(T)}\right)$$

for all $\varphi \in S(T)$. White noise $w(\omega)$ with the support in $T$ is the generalized process $w(\omega): \Omega \to \mathcal{D}'(\mathbb{R}^{r+1})$ such that: (i) $w(\varphi) = \langle w(\omega), \varphi\rangle = \langle \omega, \varphi \upharpoonright T\rangle$ (ii) $\mathbf{E}[w(\varphi)] = 0$, (iii) $\mathbf{E}[w^2(\varphi)] = \|\varphi\|^2_{L_2(T)}$.
Viewed as a Colombeau random generalized function, it has a representative (denoting on variables in $\mathbb{R}^{r+1}$ by $(x, t)$): $R_w(\varphi, x, t, \omega) = \langle \omega, \varphi(x-, t-) \upharpoonright T\rangle$, which vanishes if $t$ is less than minus the diameter of the support of $\varphi$. Therefore $w$ is a zero on $\mathbb{R}^r \times (-\infty, 0)$ in $\mathcal{G}_\Omega(\mathbb{R}^{r+1})$. Note that its variance is the Colombeau constant: $\mathbf{E}[R^2_w(\varphi, x, t, \omega)] = \int_{\mathbb{R}^r} dy \int_0^\infty |\varphi(x-y, t-s)|^2 ds$.

**Definition 2.7.** Smoothed with respect to $\mathbb{R}^r$ white noise $\left(w_\epsilon(x, t)\right)_\epsilon$ the representative $R_w(\varphi, x, t, \omega)$ with $\varphi \in A_0(\mathbb{R}^r) \times \mathcal{D}'(\mathbb{R}^r)$, such that $\varphi = \left(\varphi_\epsilon(x)\right), \epsilon \in (0,1]$ and $\varphi_\epsilon = \epsilon^{-r}\varphi\left(\frac{x}{\epsilon}\right)\delta(t)$.

**Theorem 2.1.** [25]. (**Strong Large Deviation Principle fo SPDE**) (**I**) Let $\left(u_\epsilon(x, t, \varepsilon, \eta, \omega)\right)_\epsilon, \epsilon \in (0,1]$ be solution of the Colombeau-Ito's SPDE [26]:

$$\frac{\partial\left(u_\epsilon(x,t,\varepsilon,\eta,\omega)\right)_\epsilon}{\partial t} + \Delta[\varepsilon - (1+\Delta)^2]\left(u_\epsilon(x, t, \varepsilon, \eta, \omega)\right)_\epsilon + \left(\left(F_\epsilon\left(u_\epsilon(x, t, \varepsilon, \eta, \omega)\right)\right)_\epsilon\right)\sum_{i=1}^r \delta_i \left(F_\epsilon\left(\frac{\partial u_\epsilon(x,t,\varepsilon,\eta,\omega)}{\partial x_i}\right)\right)_\epsilon + \left(f_\epsilon(x, t)\right)_\epsilon - \sqrt{\eta}\left(w_\epsilon(x, t, \omega)\right)_\epsilon = 0, \quad (2.3)$$

$$x \in \mathbb{R}^r, u_\epsilon(x, t, \varepsilon, \eta, \omega) \equiv 0, \delta_j > 0, j = 1, \ldots, r. \quad (2.4)$$

Here: (1)$\left(F_\epsilon(z)\right)_\epsilon \in G(\mathbb{R})$, $G(\mathbb{R})$ the Colombeau algebra of Colombeau generalized functions and $F_0(z) = z$.
(2) $\left(f_\epsilon(x,t)\right)_\epsilon \in \mathcal{G}(\mathbb{R}^{r+1})$.

(3) $w_\epsilon(x,t)$ is a smoothed with respect to $\mathbb{R}^r$ white noise.

**(II)** Let $\left(u_{\epsilon,\boldsymbol{n}}(x_{\boldsymbol{n}},t,\varepsilon,\eta,\omega)\right)_\epsilon, \epsilon \in (0,1]$ be solution of the Colombeau-Ito's SDE [26]:

$$\frac{d\left(u_{\epsilon,\boldsymbol{n}}(x_{\boldsymbol{n}},t,\varepsilon,\eta,\omega)\right)_\epsilon}{dt} + \Delta_{\boldsymbol{n}}[\varepsilon - (1+\Delta_{\boldsymbol{n}})^2]\left(u_{\epsilon,\boldsymbol{n}}(x_{\boldsymbol{n}},t,\varepsilon,\eta,\omega)\right)_\epsilon +$$
$$\left(\left(F_\epsilon\left(u_{\epsilon,\boldsymbol{n}}(x_{\boldsymbol{n}},t,\varepsilon,\eta,\omega)\right)\right)_\epsilon\right)\sum_{i=1}^r \delta_i \left(F_\epsilon\left(\frac{u_{\epsilon,\boldsymbol{n}+1,i}(x_{\boldsymbol{n}+1,i},t,\varepsilon,\eta,\omega) - u_{\epsilon,\boldsymbol{n}}(x_{\boldsymbol{n},i},t,\varepsilon,\eta,\omega)}{h_N}\right)\right)_\epsilon + \left(f_\epsilon(x_{\boldsymbol{n}},t)\right)_\epsilon -$$
$$\sqrt{\eta}\left(w_\epsilon(x_{\boldsymbol{n}},t,\omega)\right)_\epsilon = 0, \quad (2.5)$$

$x_{\boldsymbol{n}} \in \boldsymbol{\Theta} \subset h_N \cdot \mathbb{Z}^r$, $\boldsymbol{n} = (n_1,\dots,n_r) \in \mathbb{Z}^r, |\boldsymbol{n}| = \sum_{i=1}^r n_j, x_{\boldsymbol{n}+1,i} = \left(x_{n_1},\dots,x_{n_i} + h_N,\dots,x_{n_r}\right), -N \le n_i \le N$,(2.6)

$u_{\epsilon,\boldsymbol{n}}(x_{\boldsymbol{n}},0,\varepsilon,\eta,\omega) \equiv 0,\ x_{\boldsymbol{n}} \in \boldsymbol{\Theta}.$ (2.7)

Here Eq.(2.5)-Eq.(2.7) is obtained from Eq.(2.3)-Eq.(2.4) by spatial discretization on finite lattice $\boldsymbol{\Theta}$
$h_N \to 0$ if $N \to \infty$ and $\Delta_{\boldsymbol{n}}$-is a latticed Laplacian [31]-[33].

**(III)** Assume that Colombeau-Ito's SDE (2.5)-(2.7) is a strongly dissipative. [26].

**(IV)** Let $\Re(x,t,\varepsilon,\lambda)$ be the solutions of the linear PDE:

$$\frac{\partial\Re(x,t,\varepsilon,\lambda)}{\partial t} + \Delta[\varepsilon - (1+\Delta)^2]\Re(x,t,\varepsilon,\lambda) + \lambda\sum_{i=1}^r \delta_i \frac{\partial\Re(x,t,\varepsilon,\lambda)}{\partial x_i} - f(x,t) = 0, \lambda \in \mathbb{R}, \quad (2.8)$$

$\Re(x,0,\varepsilon,\lambda) = 0.$ (2.9)

Then

$$\liminf_{\epsilon\to 0}\mathbf{E}[|u_\epsilon(x,t,\varepsilon,\eta,\omega) - \lambda|^2] \le \Re(x,t,\varepsilon,\lambda). \quad (2.10)$$

**Proof.** The proof based on Strong large deviations principles (SLDP-Theorem) for Colombeau-Ito's solution of the Colombeau-Ito's SDE, see [26], theorem 6. By SLDP-Theorem one obtain directly the differential master equation (see [26],Eq.(90)) for Colombeau-Ito's SDE (2.5)- (2.7):

$$\frac{d\left(U_{\epsilon,\boldsymbol{n}}(x_{\boldsymbol{n}},t,\varepsilon)\right)_\epsilon}{dt} + \Delta_{\boldsymbol{n}}[\varepsilon - (1+\Delta_{\boldsymbol{n}})^2]\left(U_{\epsilon,\boldsymbol{n}}(x_{\boldsymbol{n}},t,\varepsilon)\right)_\epsilon +$$

$$+\lambda_{\boldsymbol{n}}\sum_{i=1}^r \delta_i\left(F_\epsilon\left(\frac{U_{\epsilon,\boldsymbol{n}+1,i}(x_{\boldsymbol{n}+1,i},t,\varepsilon) - U_{\epsilon,\boldsymbol{n}}(x_{\boldsymbol{n},i},t,\varepsilon)+}{h_N}\right)\right)_\epsilon + \left(f_\epsilon(x_{\boldsymbol{n}},t)\right)_\epsilon + o(\epsilon) = 0, \quad (2.11)$$

$U_{\epsilon,\boldsymbol{n}}(x_{\boldsymbol{n}},0,\varepsilon) = -\lambda_{\boldsymbol{n}}.$ (2.12)

We set now $\lambda_{\boldsymbol{n}} \equiv \lambda \in \mathbb{R}$. Then from Eq. (2.13)-Eq. (2.14) we obtain

$$\frac{d\left(U_{\epsilon,\boldsymbol{n}}(x_{\boldsymbol{n}},t,\varepsilon,\lambda)\right)_{\epsilon}}{dt} + \Delta_{\boldsymbol{n}}[\varepsilon - (1+\Delta_{\boldsymbol{n}})^2]\left(U_{\epsilon,\boldsymbol{n}}(x_{\boldsymbol{n}},t,\varepsilon,\lambda)\right)_{\epsilon} +$$

$$+\lambda \sum_{i=1}^{r} \delta_i \left(F_{\epsilon}\left(\frac{U_{\epsilon,\boldsymbol{n}+1,i}(x_{\boldsymbol{n}+1,i},t,\varepsilon,\lambda) - U_{\epsilon,\boldsymbol{n}}(x_{\boldsymbol{n},i},t,\varepsilon,\lambda)}{h_N}\right)\right)_{\epsilon} + \left(f_{\epsilon}(x_{\boldsymbol{n}},t)\right)_{\epsilon} + O(\epsilon) = 0, \quad (2.13)$$

$$U_{\epsilon,\boldsymbol{n}}(x_{\boldsymbol{n}},0,\varepsilon,\lambda) = -\lambda. \quad (2.14)$$

From Eq. (2.5)-Eq. (2.7) and Eq. (2.13)-Eq. (2.14) by SLDP-Theorem (see see [26], inequality (89)) we obtain the inequality

$$\operatorname{liminf}_{\epsilon\to 0}\mathbf{E}\left[\left|u_{\epsilon,\boldsymbol{n}}(x_{\boldsymbol{n}},t,\varepsilon,\eta,\omega) - \lambda\right|^2\right] \leq U_{\epsilon,\boldsymbol{n}}(x_{\boldsymbol{n}},t,\varepsilon,\lambda), \epsilon \in (0,1]. \quad (2.15)$$

Let us consider now the identity

$$|u_{\epsilon}(x,t,\varepsilon,\eta,\omega) - \lambda|^2 = \left|\left[u_{\epsilon}(x,t,\varepsilon,\eta,\omega) - u_{\epsilon,\boldsymbol{n}}(x_{\boldsymbol{n}},t,\varepsilon,\eta,\omega)\right] + \left[u_{\epsilon,\boldsymbol{n}}(x_{\boldsymbol{n}},t,\varepsilon,\eta,\omega) - \lambda\right]\right|^2. \quad (2.16)$$

From the identity (2.16) by the triangle inequality we obtain the inequality

$$|u_{\epsilon}(x,t,\varepsilon,\eta,\omega) - \lambda|^2 \leq \left|u_{\epsilon}(x,t,\varepsilon,\eta,\omega) - u_{\epsilon,\boldsymbol{n}}(x_{\boldsymbol{n}},t,\varepsilon,\eta,\omega)\right|^2 + \left|u_{\epsilon,\boldsymbol{n}}(x_{\boldsymbol{n}},t,\varepsilon,\eta,\omega) - \lambda\right|^2. \quad (2.17)$$

From the identity (2.17) by integration we obtain the inequality

$$\mathbf{E}|u_{\epsilon}(x,t,\varepsilon,\eta,\omega) - \lambda|^2 \leq$$

$$\leq \mathbf{E}\left|u_{\epsilon}(x,t,\varepsilon,\eta,\omega) - u_{\epsilon,\boldsymbol{n}}(x_{\boldsymbol{n}},t,\varepsilon,\eta,\omega)\right|^2 + \left|u_{\epsilon,\boldsymbol{n}}(x_{\boldsymbol{n}},t,\varepsilon,\eta,\omega) - \lambda\right|^2. \quad (2.18)$$

From the identity (2.18) by the identity (2.15) for all $\epsilon \in (0,1]$ we obtain the inequality

$$\mathbf{E}|u_{\epsilon}(x,t,\varepsilon,\eta,\omega) - \lambda|^2 \leq$$

$$\leq \mathbf{E}\left|u_{\epsilon}(x,t,\varepsilon,\eta,\omega) - u_{\epsilon,\boldsymbol{n}}(x_{\boldsymbol{n}},t,\varepsilon,\eta,\omega)\right|^2 + U_{\epsilon,\boldsymbol{n}}(x_{\boldsymbol{n}},t,\varepsilon,\lambda). \quad (2.19)$$

In the limit $N \to \infty$ from the inequality we obtain the inequality

$$\mathbf{E}|u_{\epsilon}(x,t,\varepsilon,\eta,\omega) - \lambda|^2 \leq$$

$$\leq \operatorname{limsup}_{N\to\infty}\mathbf{E}\left|u_{\epsilon}(x,t,\varepsilon,\eta,\omega) - u_{\epsilon,\boldsymbol{n}}(x_{\boldsymbol{n}},t,\varepsilon,\eta,\omega)\right|^2 + \operatorname{limsup}_{N\to\infty}U_{\epsilon,\boldsymbol{n}}(x_{\boldsymbol{n}},t,\varepsilon,\lambda). \quad (2.20)$$

We note that

$$\operatorname{limsup}_{N\to\infty}\mathbf{E}\left|u_{\epsilon}(x,t,\varepsilon,\eta,\omega) - u_{\epsilon,\boldsymbol{n}}(x_{\boldsymbol{n}},t,\varepsilon,\eta,\omega)\right|^2 = 0. \quad (2.21)$$

Therefore from (2.20) and (2.21) we obtain the inequality

$$\mathbf{E}|u_{\epsilon}(x,t,\varepsilon,\eta,\omega) - \lambda|^2 \leq \operatorname{limsup}_{N\to\infty}U_{\epsilon,\boldsymbol{n}}(x_{\boldsymbol{n}},t,\varepsilon,\lambda) \quad (2.22)$$

In the limit $N \to \infty$ from Eq.(2.13)-Eq.(2.14) for any fixed $\epsilon \neq 0, \epsilon \ll 1$, we obtain the differential master equation for Colombeau-Ito's SPDE (2.3)-(2.4)

$$\frac{d\left(U_\epsilon(x,t,\varepsilon,\lambda)\right)_\epsilon}{dt} + \Delta[\varepsilon - (1+\Delta)^2]\left(U_\epsilon(x,t,\varepsilon,\lambda)\right)_\epsilon +$$

$$+\lambda \sum_{i=1}^{r} \delta_i \left( F_\epsilon\left(\frac{\partial U_\epsilon(x,t,\varepsilon,\lambda)}{\partial x_i}\right)\right)_\epsilon + \left(f_\epsilon(x,t)\right)_\epsilon + O(\epsilon) = 0, \quad (2.23)$$

$$U_\epsilon(x,t,\varepsilon,\lambda) = -\lambda. \quad (2.24)$$

Therefore from the inequality (2.22) follows the inequality

$$\mathbf{E}|u_\epsilon(x,t,\varepsilon,\eta,\omega) - \lambda|^2 \leq U_\epsilon(x,t,\varepsilon,\lambda). \quad (2.25)$$

In the limit $\epsilon \to 0$ from differential equation (2.23)-(2.24) we obtain the differential equation (2.8)-(2.9) and it is easy to see that

$$\lim_{\epsilon\to 0} U_\epsilon(x,t,\varepsilon,\lambda) = \Re(x,t,\varepsilon,\lambda). \quad (2.26)$$

From the inequality (2.25) one obtain the inequality

$$\operatorname{liminf}_{\epsilon\to 0}\mathbf{E}|u_\epsilon(x,t,\varepsilon,\eta,\omega) - \lambda|^2 \leq \lim_{\epsilon\to 0} U_\epsilon(x,t,\varepsilon,\lambda) = \Re(x,t,\varepsilon,\lambda). \quad (2.27)$$

From the inequality (2.27) and Eq.(2.26) finally we obtain the inequality

$$\operatorname{liminf}_{\epsilon\to 0}\mathbf{E}|u_\epsilon(x,t,\varepsilon,\eta,\omega) - \lambda|^2 \leq \Re(x,t,\varepsilon,\lambda). \quad (2.28)$$

The inequality (2.28) finalized the proof.

**Definition 2.7.** (**The Differential Master Equation**) The linear PDE:

$$\frac{\partial \Re(x,t,\varepsilon,\lambda)}{\partial t} + \Delta[\varepsilon - (1+\Delta)^2]\Re(x,t,\varepsilon,\lambda) + \lambda \sum_{i=1}^{r} \delta_i \frac{\partial \Re(x,t,\varepsilon,\lambda)}{\partial x_i} - f(x,t) = 0, \lambda \in \mathbb{R}, \quad (2.29)$$

$\Re(x,0,\varepsilon,\lambda) = 0$ (2.30),

we will call as the differential master equation.

**Definition 2.8.** (**The Transcendental Master Equation**) The transcendental equation

$$\Re\left(x,t,\varepsilon,\lambda(x,t,\varepsilon)\right) = 0, \quad (2.31)$$

we will call as the transcendental master equation.

**Remark2.2.** We note that concrete structure of the Nikolaevskii chaos is determined by the solutions $\lambda(x,t,\varepsilon)$ variety by transcendental master equation (2.31). Master equation (2.31) is determines by the only way some many-valued function $\lambda(x,t,\varepsilon)$ which is the main constructive object, determining the characteristics of quantum chaos in the corresponding model of Euclidian quantum field theory.

## 3.Criterion of the existence quantum chaos in Euclidian quantum N-model.

**Definition3.1.**Let $u_\eta(x,t,\varepsilon,\omega)$be the solution of the Eq.(2.1). Assume that for almost all points$(x,t) \in \mathbb{R}^r \times \mathbb{R}_+$(in the sense of Lebesgue–measureon $\mathbb{R}^r \times \mathbb{R}_+$), there exist a function $u(x,t)$ such that

$$\lim_{\eta\to 0}\mathbf{E}\left[\left(u_\eta(x,t,\varepsilon,\omega) - u(x,t)\right)^2\right] = 0. \qquad (3.1)$$

Then we will say that a function $u(x,t)$is a quasi-determined solution (QD-solution of the Eq.(2.).

**Definition3.2.** Assume that there exist a set$\mathfrak{H} \subset \mathbb{R}^r \times \mathbb{R}_+$that is positive Lebesgue–measure, i.e.,$\mu(\mathfrak{H}) > 0$ and

$$\forall(x,t)\{(x,t) \in \mathfrak{H} \to \neg\exists\lim_{\eta\to 0}\mathbf{E}[u_\eta^2(x,t,\varepsilon,\omega)]\},(3.2)$$

i.e., $(x,t) \in \mathfrak{H}$ imply that the limit: $\lim_{\eta\to 0}\mathbf{E}[u_\eta^2(x,t,\varepsilon,\omega)]$does not exist.

Then we will say that Euclidian quantum N-model has the quasi-determined Euclidian quantum chaos (**QD**-quantum chaos).

**Definition3.3.**For each point$(x,t) \in \mathbb{R}^r \times \mathbb{R}_+$we define a set $\{\widetilde{\mathfrak{R}}(x,t,\varepsilon)\} \subset \mathbb{R}$ by the condition:

$$\forall\lambda[\lambda \in \{\widetilde{\mathfrak{R}}(x,t,\varepsilon)\} \Leftrightarrow \mathfrak{R}(x,t,\varepsilon,\lambda) = 0\ ].(3.3)$$

**Definition3.4.**Assume that Euclidian quantum N-model(2.1) has the Euclidian **QD**-quantum chaos. For each point $(x,t) \in \mathbb{R}^r \times \mathbb{R}_+$ we define a set-valued function$\widetilde{\mathfrak{R}}(x,t)$: $\mathbb{R}^r \times \mathbb{R}_+ \to 2^{\mathbb{R}}$ by the condition:

$$\widetilde{\mathfrak{R}}(x,t,\varepsilon) = \{\widetilde{\mathfrak{R}}(x,t,\varepsilon)\}(3.4)$$

We will say that the set -valued function$\widetilde{\mathfrak{R}}(x,t,\varepsilon)$is a quasi-determined chaotic solution(**QD**-chaotic solution)of the quantum N-model.

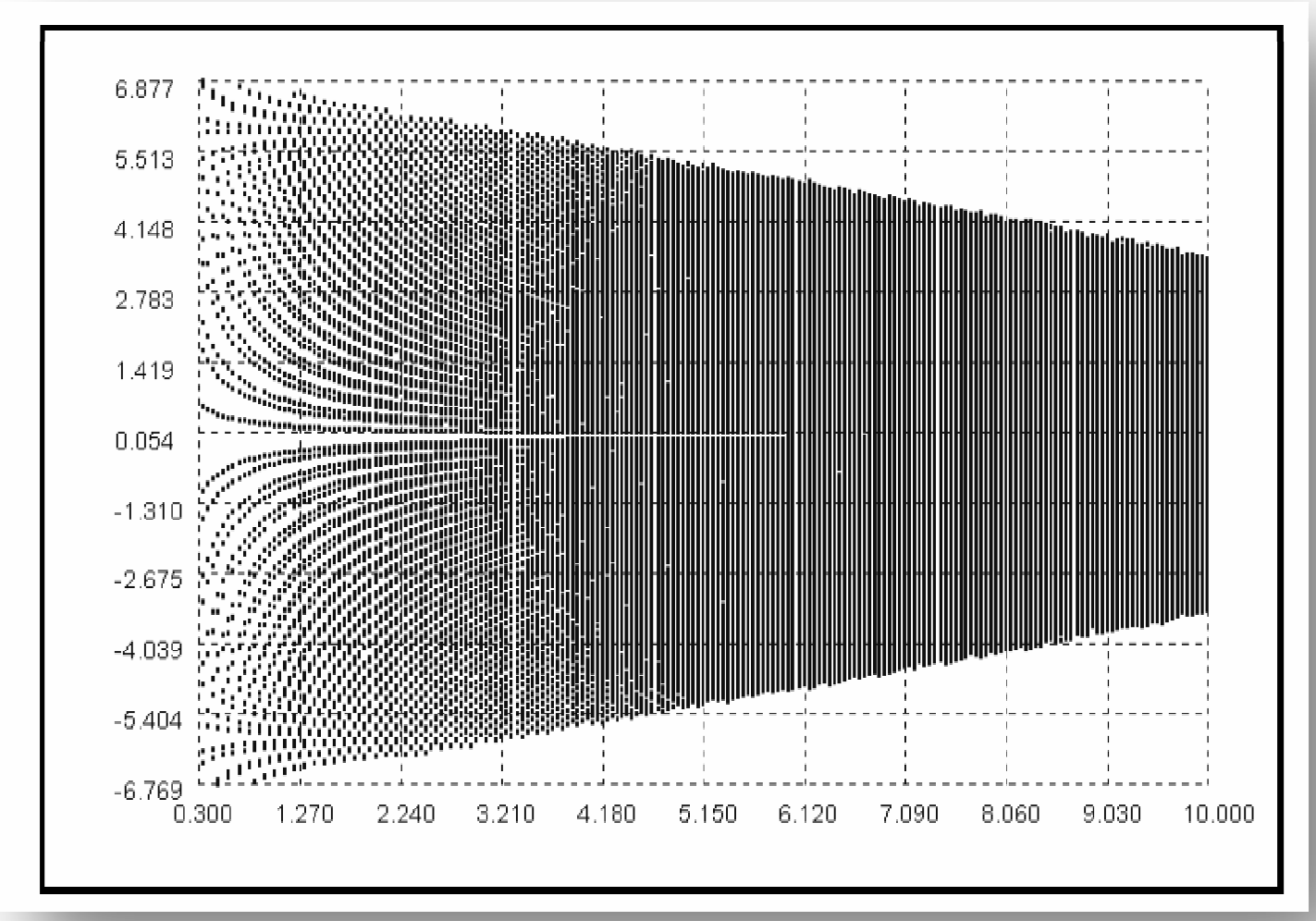

**Pic.3.1.**Evolution of **QD**-chaotic solution$\widetilde{\Re}(x,t,\varepsilon)$in time $t \in [0,10]$ at point $x = 3. t \in [0,10], \varepsilon = -10^{-2}, \sigma = 10^3, p = 1.1$.

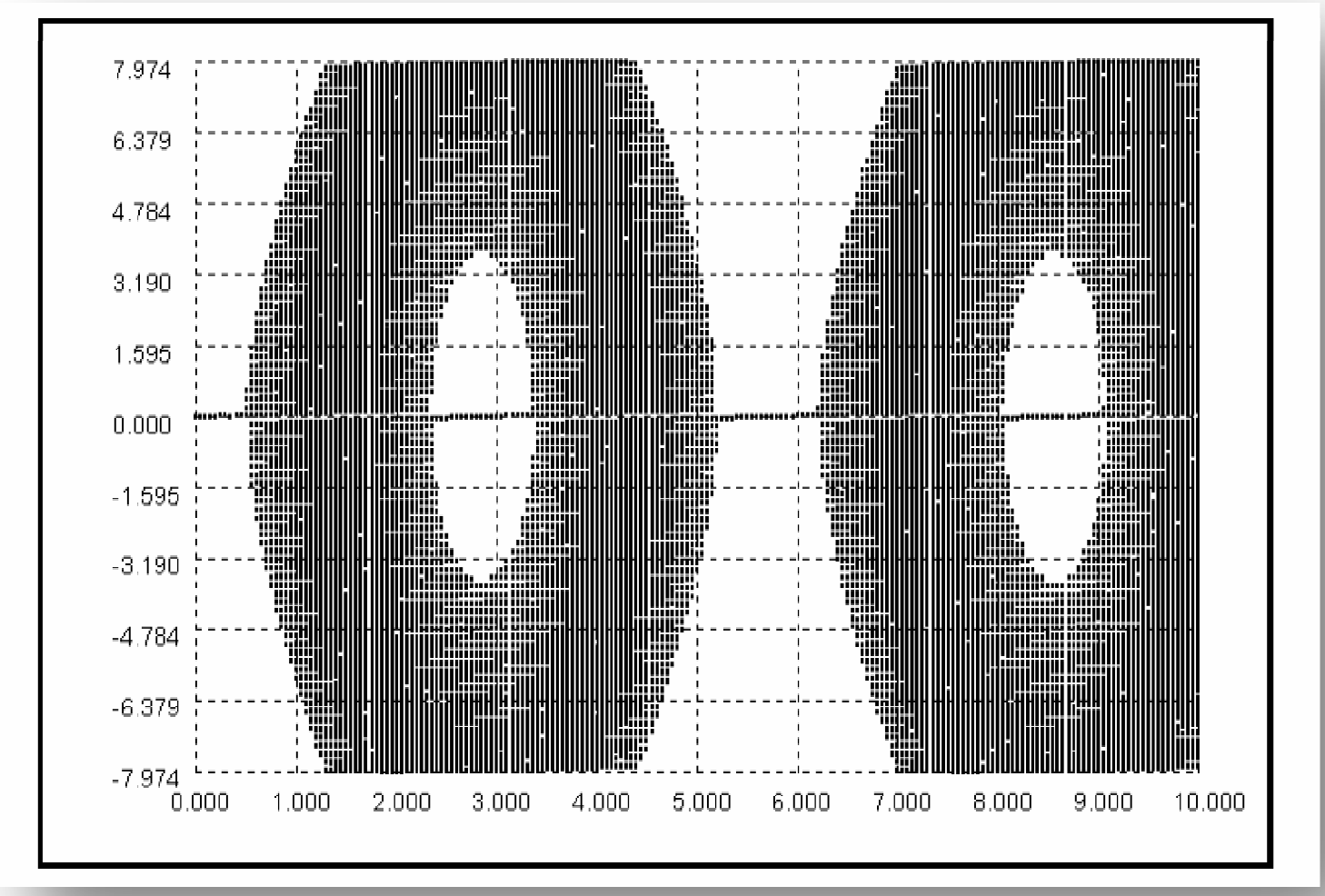

**Pic.3.2.**The spatial structure of **QD**-chaotic solution$\widetilde{\Re}(x,t,\varepsilon)$ at instant $t = 3, \varepsilon = -10^{-2}, \sigma = 10^3, p = 1.1$.

**Theorem3.1.**Assume that$f(x,t) = \sigma \sin(p \cdot x)$Then for all values of parameters$r, \varepsilon, \sigma, \delta_j, j = 1, \dots, r$ such that$r \in \mathbb{N}, \delta_j \in \mathbb{R}_+, j = 1, \dots, r, \varepsilon \in [-1,1], p \in \mathbb{R}^r, \sigma \neq 0$, quantum N-model (2.1) has the **QD**-chaotic solutions.

**Definition3.5.**For each point $(x,t) \in \mathbb{R}^r \times \mathbb{R}_+$ we define the functions such that:

(i) $u_+(x,t,\varepsilon) = \mathrm{limsup}_{\eta \to 0} \mathbf{E}[u_\eta(x,t,\varepsilon,\omega)]$,

(ii) $u_-(x,t,\varepsilon) = \mathrm{liminf}_{\eta \to 0} \mathbf{E}[u_\eta(x,t,\varepsilon,\omega)]$,

(iii) $u_w(x,t,\varepsilon) = u_+(x,t,\varepsilon,\omega) - u_-(x,t,\varepsilon,\omega)$.

**Definition3.7.**

(i) Function $u_+(x,t,\varepsilon)$is called upper bound of the **QD**-quantum chaos at point $(x,t)$.
(ii) Function $u_-(x,t,\varepsilon)$is called lower bound of the **QD**-quantum chaos at point $(x,t)$.
(iii) Function $u_w(x,t,\varepsilon)$is called width of the**QD**-quantum chaos at point$(x,t)$.

**Definition3.8.** Assume now that

$\limsup_{t\to\infty} u_w(x,t,\varepsilon) = u_w(x,\varepsilon) < \infty$.(3.5)

Then we will say thatEuclidian quantum N-model has **QD**-quantum chaos of the asymptotically finite width at point$x \in \mathbb{R}^r$.

**Definition3.9.**Assume now that

$\limsup_{t\to\infty} u_w(x,t,\varepsilon) = u_w(x,\varepsilon) = \infty$. (3.6)

Then we will say that Euclidian quantum N-model has **QD**-quantum chaos of the asymptotically infinite width at point $x \in \mathbb{R}^r$.

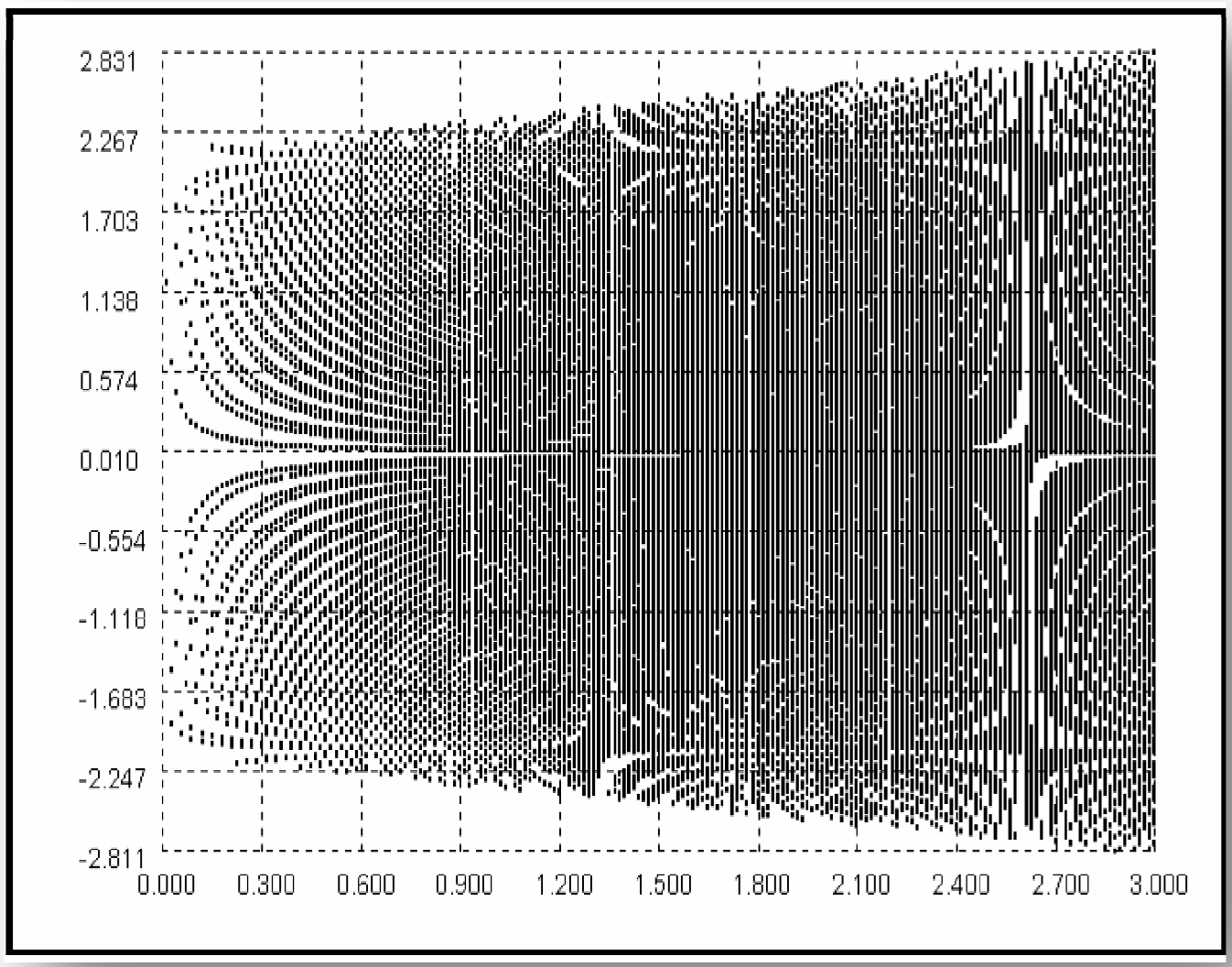

**Pic.3.3.** The**QD**-quantum chaos of the asymptotically infinite width at point $x = 3.\varepsilon = 0.1, \delta = 10, \sigma = 10^3, p = 1$.

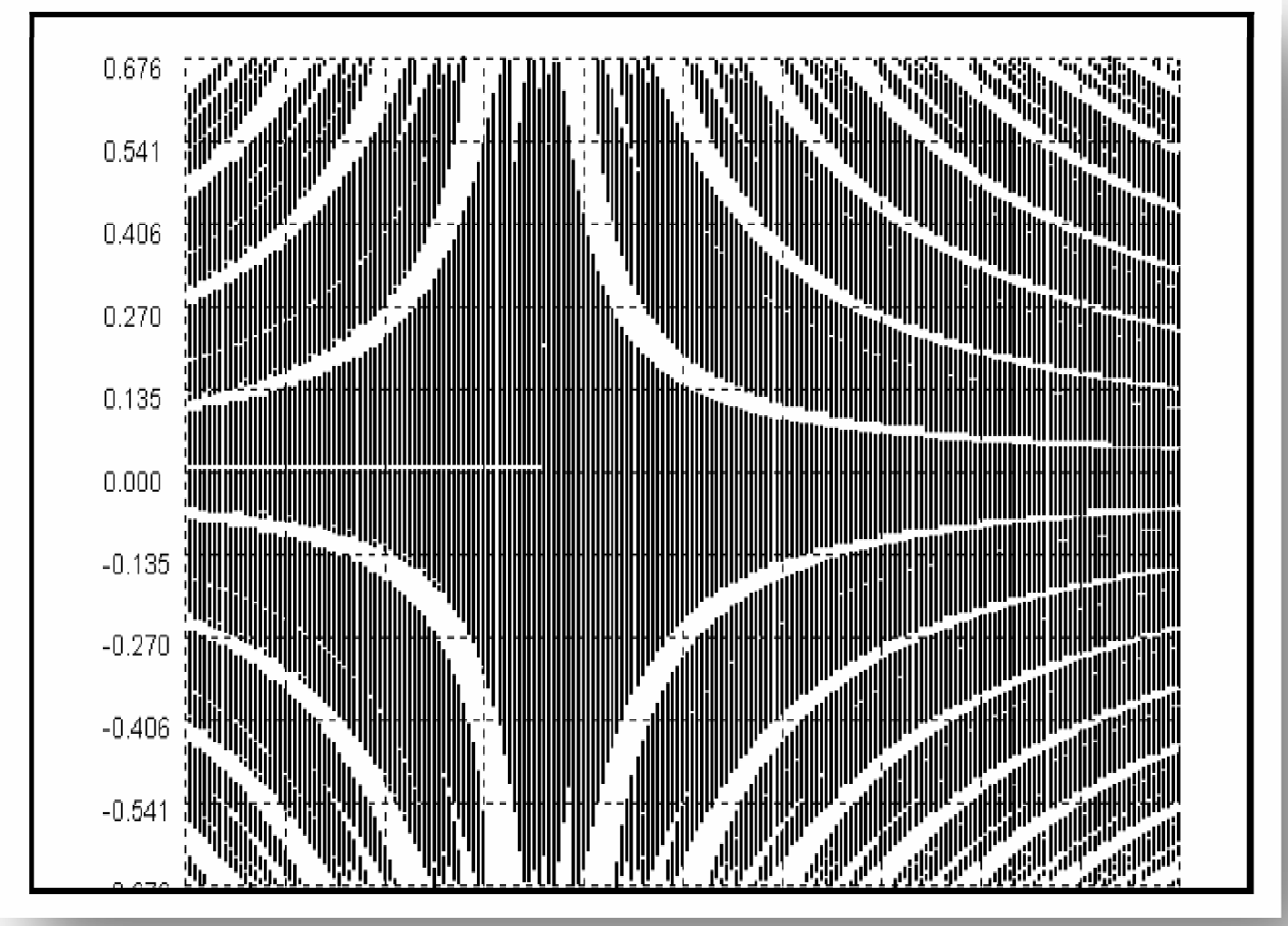

**Pic.3.4.**The fine structure of the **QD**-quantum chaos of the asymptotically infinite width at point $x = 3, \varepsilon = 0, \delta = 10,$
$\sigma = 10^4, p = 1, t \in [10^4, 10^4 + 10^{-1}], \lambda \in [-0.676, 0.676]$.

**Definition3.10**. For each point $(x,t) \in \mathbb{R}^r \times \mathbb{R}_+$ we define the functions such that:

(i) $\widetilde{\Re}_+(x,t,\varepsilon) = \sup\{\widetilde{\Re}(x,t,\varepsilon)\}$,
(ii) $\widetilde{\Re}_-(x,t,\varepsilon) = \inf\{\widetilde{\Re}(x,t,\varepsilon)\}$,
(iii) $\widetilde{\Re}_w(x,t,\varepsilon) = \widetilde{\Re}_+(x,t,\varepsilon) - \widetilde{\Re}_-(x,t,\varepsilon)$.

**Theorem3.2.** For each point $(x,t) \in \mathbb{R}^r \times \mathbb{R}_+$is satisfiedtheinequality

$\widetilde{\Re}_w(x,t,\varepsilon) \leq u_w(x,t,\varepsilon)$.(3.7)

**Proof.** Immediately follows by Theorem2.1 and Definitions 3.5, 3.10.

**Theorem3.3. (Criterion of QD-quantum chaos in Euclidian quantum N-model)**
Assume that
$\mathrm{mes}\{(x,t)|\widetilde{\Re}_w(x,t,\varepsilon) > 0\} > 0$.(3.8)

Then Euclidian quantum N-model has **QD**-quantum chaos.
**Proof.** Immediately follows by the inequality (3.7)and Definition3.2.

## 4. Quasi-determined quantum chaos and physical turbulence nature.

In generally accepted at the present time hypothesis what physical turbulence in the dynamical systems with an infinite number of degrees of freedom really is, the physical turbulence is associated with a strange attractors, on which the phase trajectories of dynamical system reveal the known properties of stochasticity: a very high dependence on the initial conditions, which is associated with exponential dispersion of the initially close trajectories and brings to their non-reproduction; everywhere the density on the attractor almost of all the trajectories a very fast decrease of local auto-correlation function[2]-[9]

$\Phi(x,\tau) = \langle \tilde{u}(x,t)\tilde{u}(x,\tau+t)\rangle$,(4.1)

Here

$$\tilde{u}(x,t) = u(x,t) - \langle u(x,t)\rangle, \langle f(t)\rangle = \lim_{T\to\infty}\langle f(t)\rangle_T, \langle f(t)\rangle_T = \frac{1}{T}\int_0^T f(t).$$

In contrast with canonical numerical simulation, by using Theorem2.1 it is possible to study non-perturbativelythe influence of thermal additive fluctuations on classical dynamics, which in the consideredcase is described by equation (4.1).

The physical nature of quasi-determined chaos is simple and mathematically is associated with discontinuously of the trajectories of the stochastic process$u_\eta(x,t,\varepsilon,\omega)$on parameter $\eta$.

In order to obtain the characteristics of this turbulence, which is a very similarly tolocal auto-correlation function (3.1) we define bellow some appropriate functions.

**Definition 4.1.**The numbering function$N(t,x)$ of quantum chaos in Euclidian quantum N-modelis defined by

$N(x,t) = \mathrm{card}\{\widetilde{\Re}(x,t)\}$.(4.2)

Here by $\mathrm{card}\{X\}$ we denote the cardinality of a finite set$X$,i.e., the number of its elements.

**Definition 4.2.**Assume now that a set $\{\widetilde{\Re}(x,t)\}$is ordered be increase of its elements. We introduce thefunction $\widetilde{\Re}_i(x,t), i = 1,\dots,N(x,t)$which value at point$(x,t)$, equals the $i$-th element of a set $\{\widetilde{\Re}(x,t)\}$.

**Definition 3.3.**The mean value function$\overline{u}(x,t)$ ofthe chaotic solution $\widetilde{\Re}(x,t)$at point $(x,t)$is defined by

$\overline{u}(x,t) = (N(x,t))^{-1}\sum_{i=1}^{N(x,t)}\widetilde{\Re}_i(x,t)$.(4.3)

**Definition 3.4.**The turbulent pulsations function $u^*(x,t)$ of the chaotic solution $\widetilde{\Re}(x,t)$at point $(x,t)$is defined by

$u^*(x,t) = \sqrt{(N(x,t))^{-1}\sum_{i=1}^{N(x,t)}|\widetilde{\Re}_i(x,t) - \overline{u}(x,t)|}$.(4.4)

**Definition3.5.**The local auto-correlation function is defined by

$\Phi(x,\tau) = \lim_{T\to\infty}\langle \tilde{u}(x,\tau)\tilde{u}(x,\tau+t)\rangle_T = \lim_{T\to\infty}\frac{1}{T}\int_0^T \tilde{u}(x,t)\tilde{u}(x,\tau+t)dt$,(4.5)

$\tilde{u}(x,t) = \overline{u}(x,t) - \breve{u}(x), \breve{u}(x) = \lim_{T\to\infty}\frac{1}{T}\int_0^T \overline{u}(x,t)dt$.(4.6)

**Definition 3.5.**The normalized local auto-correlation function is defined by

$\Phi_n(x,\tau) = \frac{\Phi(x,\tau)}{\Phi(x,0)}$.(4.7)

Let us consider now 1DEuclidian quantum N-model corresponding to classical dynamics

$$\frac{\partial^2}{\partial x^2}\left[\varepsilon-\left(1+\frac{\partial^2}{\partial x^2}\right)^2\right]u(x,\varepsilon)+\delta\frac{\partial u(x,\varepsilon)}{\partial x}u(x,\varepsilon)-\sigma\sin(p\cdot x)=0,(4.8)$$

Corresponding Langevin equation are [34]-[35]:

$$\frac{\partial u_\eta(x,t,\varepsilon)}{\partial t}+\Delta[\varepsilon-(1+\Delta)^2]u_\eta(x,t,\varepsilon)+\delta\frac{\partial u_\eta(x,t,\varepsilon)}{\partial x}u_\eta(x,t,\varepsilon)-$$

$$-\sigma\sin(px)=\sqrt{\eta}w(x,t),\delta>0\Delta=\frac{\partial^2}{\partial x^2},(4.9)$$

$$u_\eta(x,0,\varepsilon)=0,w(x,t)=\frac{\partial^2 W(x,t)}{\partial x\partial t}.(4.10)$$

Corresponding differential master equation are

$$\frac{\partial\Re(x,t,\varepsilon,\lambda)}{\partial t}+\Delta[\varepsilon-(1+\Delta)^2]\Re(x,t,\varepsilon,\lambda)+\lambda\delta\frac{\partial\Re(x,t,\varepsilon,\lambda)}{\partial x}-\sigma\sin(px)=0,(4.11)$$

$$\Re(x,0,\varepsilon,\lambda)=-\lambda.(4.12)$$

Corresponding transcendental master equation (2.29)-(2.30) are

$$\frac{\{\cos(p\cdot x)-\exp[t\cdot\chi(p)]\cos[p(x-\lambda\cdot\delta\cdot t)]\}\cdot\lambda\cdot\delta\cdot p}{\chi^2(p)+\lambda^2\cdot\delta^2\cdot p^2}+\frac{\{\sin(p\cdot x)-\exp[t\cdot\chi(p)]\sin[p(x-\lambda\cdot\delta\cdot t)]\}\cdot\chi(p)}{\chi^2(p)+\lambda^2\cdot\delta^2\cdot p^2}+\frac{\lambda}{\sigma}=0,(4.13)$$

$$\chi(p)=p^2[\varepsilon-(p^2-1)^2].(4.14)$$

We assume now that $\chi(p)=0$.Then from Eq.(4.13) for all $t\in[0,\infty)$ we obtain

$$\frac{\{\cos(p\cdot x)-\cos[p(x-\lambda\cdot\delta\cdot t)]\}\cdot\lambda\cdot\delta\cdot p}{\lambda^2\cdot\delta^2\cdot p^2}+\frac{\lambda}{\sigma}=0,\quad \text{or} \qquad (4.14)$$

$$\{\cos(p\cdot x)-\cos[p(x-\lambda\cdot\delta\cdot t)]\}\cdot\sigma\cdot\delta^{-1}\cdot p^{-1}+\lambda^2=0.(4.15)$$

The result of calculation using transcendental master equation (4.15) the corresponding function $\widetilde{\Re}(x,t,\varepsilon)$ is presented by Pic.4.1 and Pic.4.2.

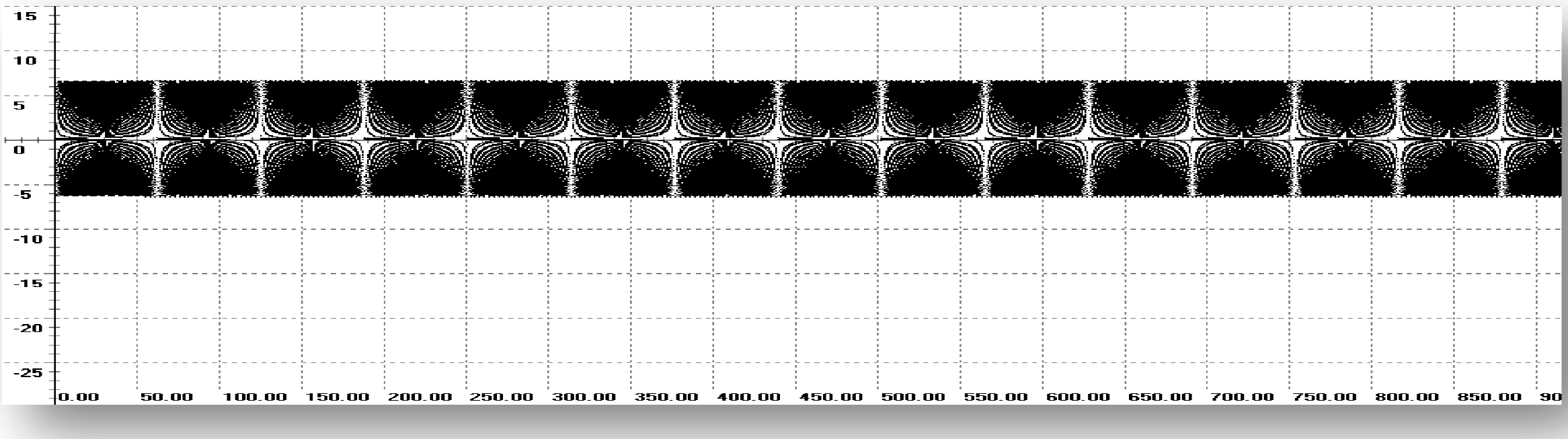

**Pic.4.1.** Evolution of **QD**-chaotic solution $\widetilde{\Re}(10^3,t,\varepsilon)$ in time $t\in[0,10^3]$, $\Delta t=0.1,\varepsilon=0,p=1,$ $\sigma=10^2,\delta=1,,\Delta\lambda=0.01.$

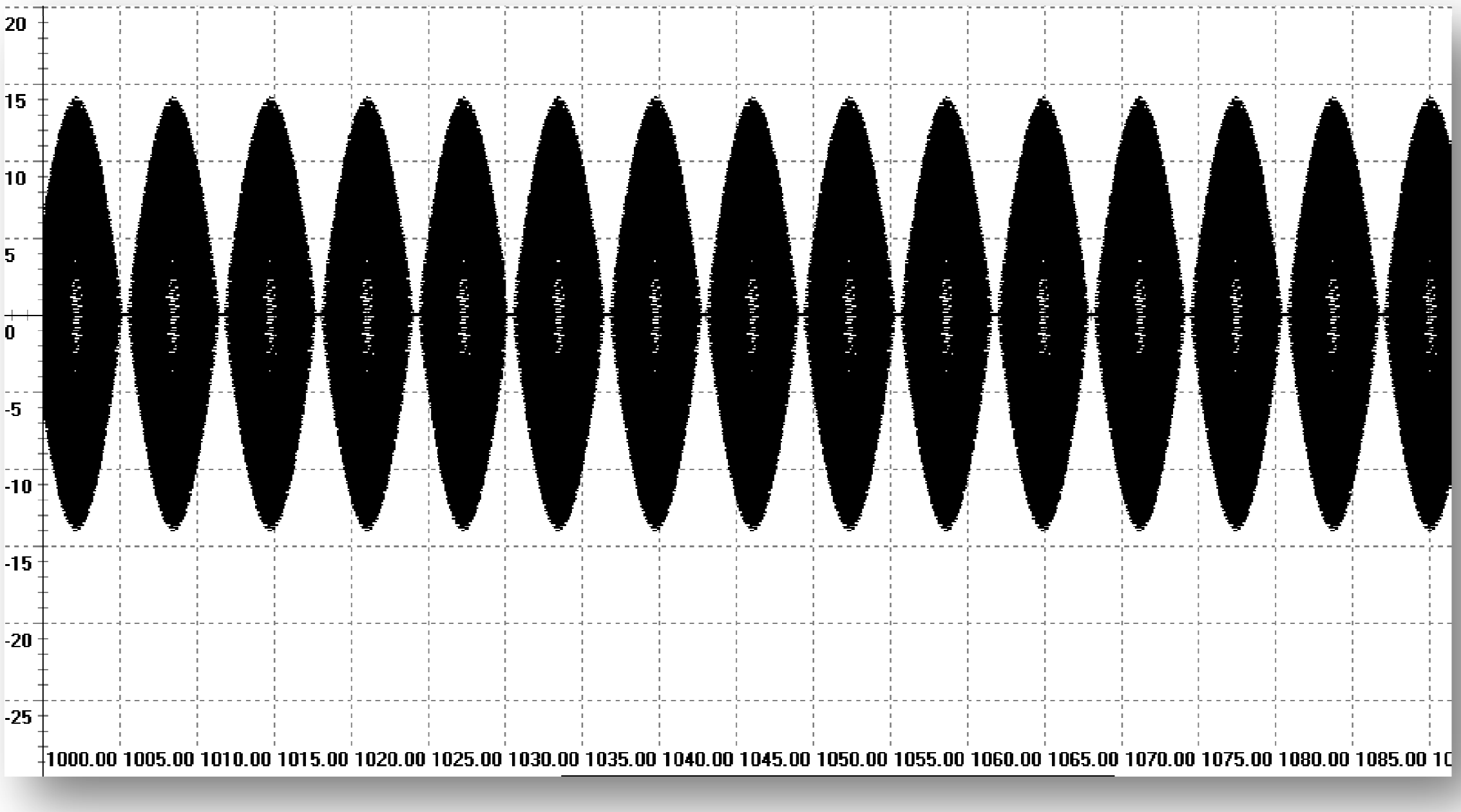

**Pic.4.2.** The spatial structure of **QD**-chaotic solution $\widetilde{\Re}(x,t,\varepsilon)$ at instant $t = 10^3, \varepsilon = 0, p = 1, \sigma = 10^2$, $\delta = 1,\ \Delta x = 0.1, \Delta\lambda = 0.01$.

The result of calculation using master equation(4.13) the corresponding function $\widetilde{\Re}(x,t,\varepsilon)$ is presented by Pic.4.3 and Pic.4.4.

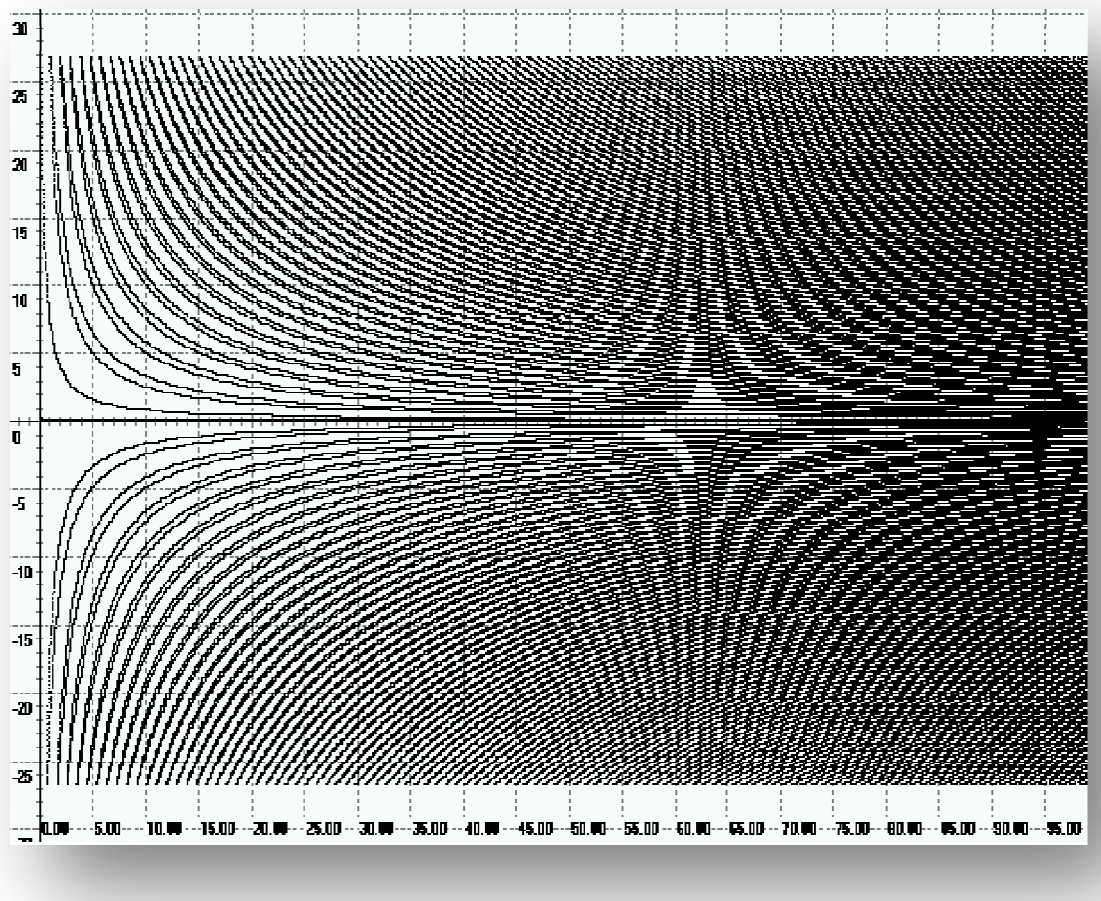

**Pic.4.3.**The development of temporal chaotic regime of1D Euclidian quantum **N**-model at point $x = 1, t \in [0, 10^2]$. $\varepsilon = 10^{-7}, \sigma = 10^2, \delta = 1, p = 1$.

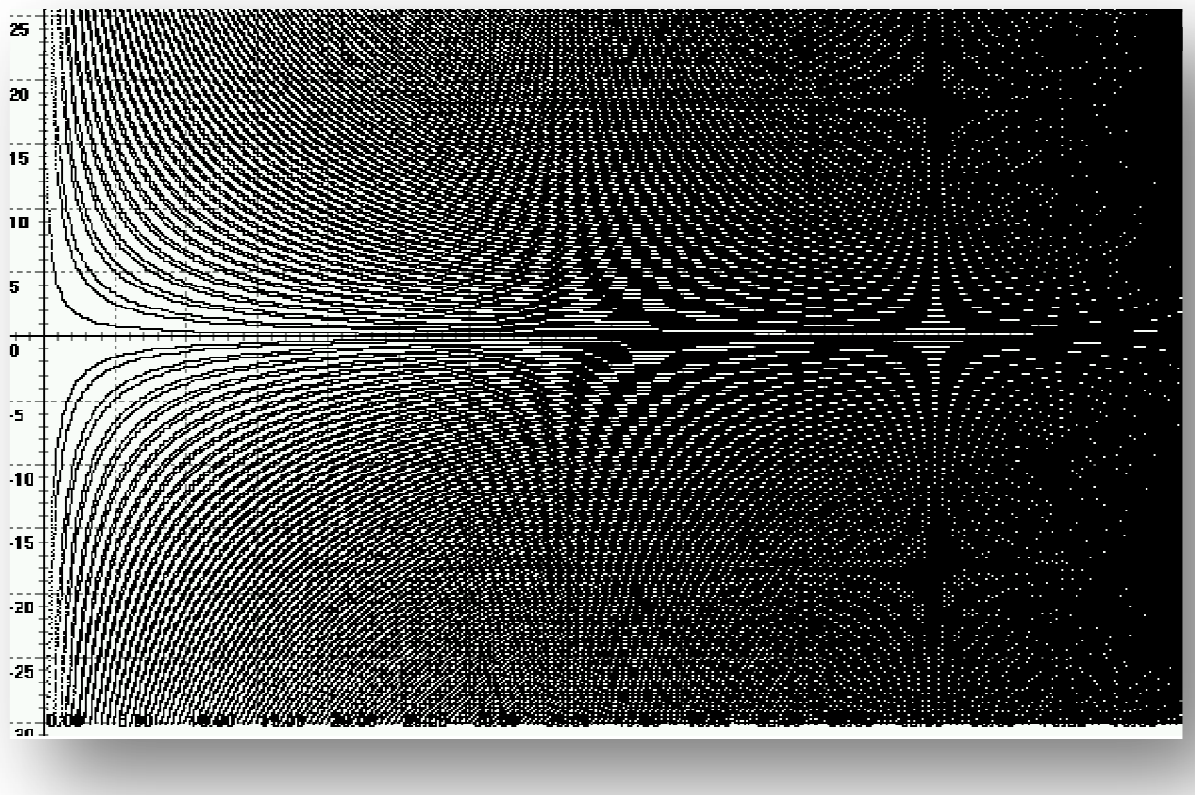

**Pic.4.4.**The developmentof temporal chaotic regimeof 1D Euclidian quantum N-model at point $x = 1, t \in [0, 10^2]$, $\varepsilon = 10^{-7}, \sigma = 5 \cdot 10^5, \delta = 1, p = 1$.

Let us calculate now corresponding normalized local auto-correlation function $\Phi_n(x, \tau)$. The result of calculation using Eq.(4.7)-Eq.(4.7) is presented by Pic.4.5 and Pic.4.6.

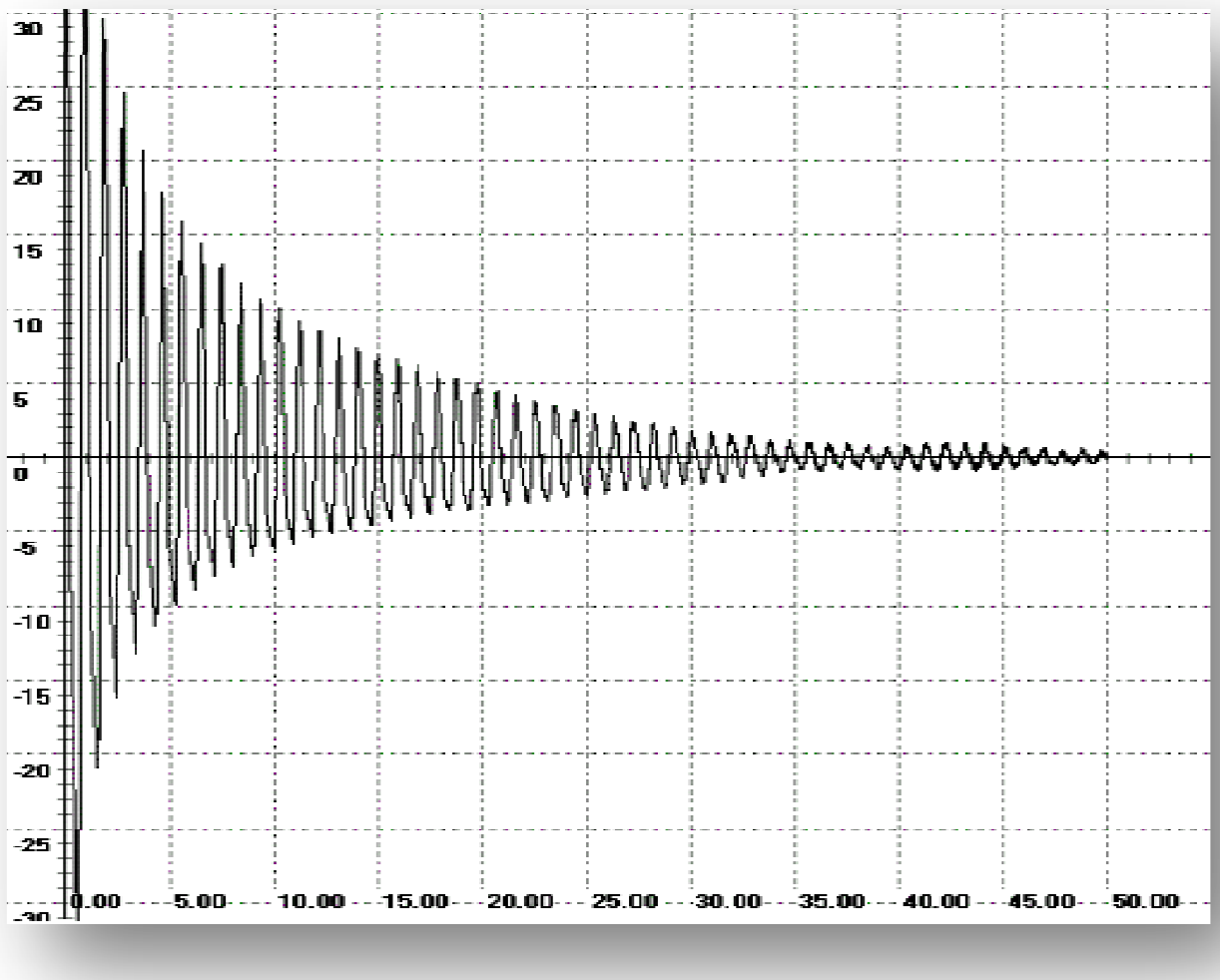

**Pic.4.5.** Normalized local auto-correlation function $\Phi_n(1, \tau)$ $t \in [0,50]$, $\varepsilon = 10^{-7}, \sigma = 10^2, \delta = 1, p = 1$.

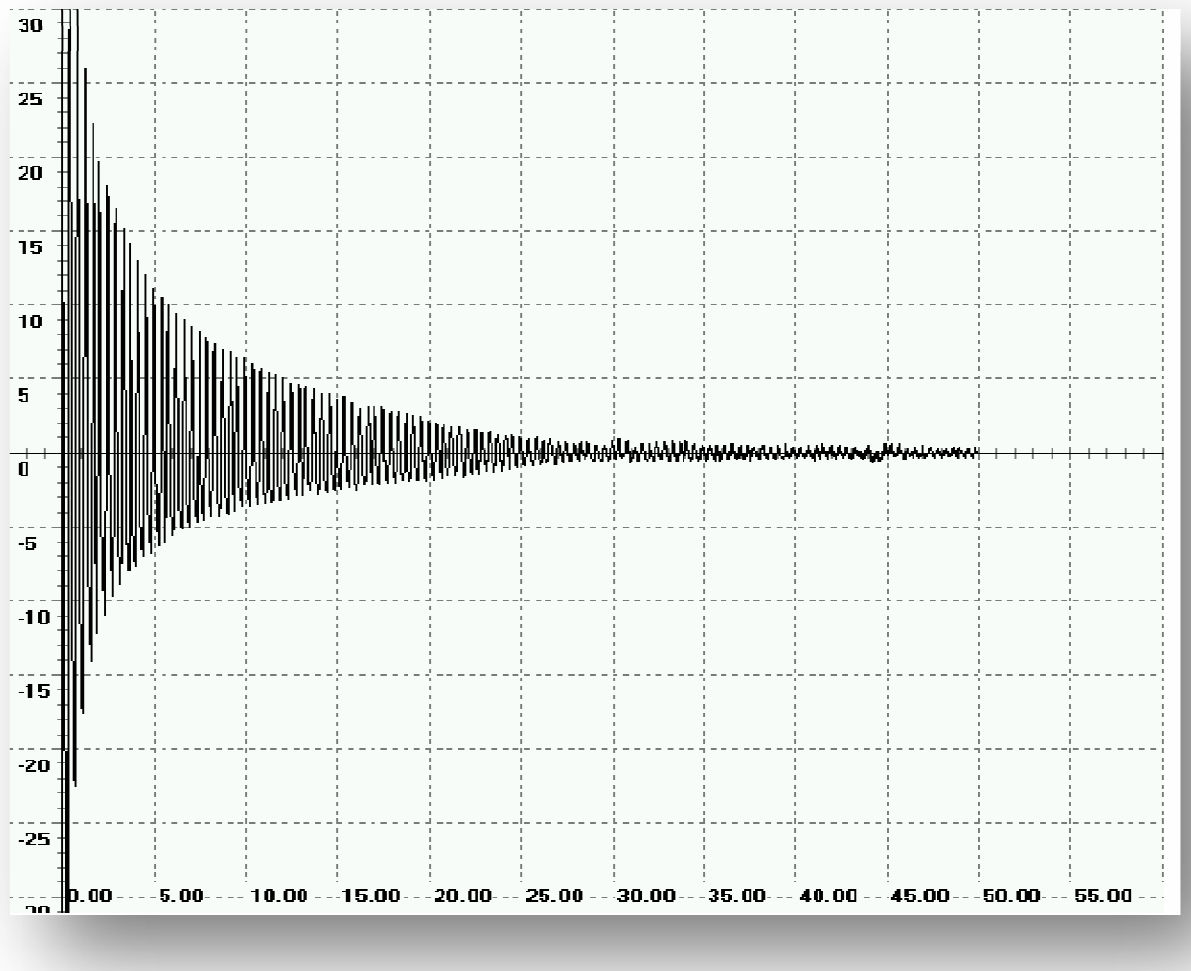

**Pic.4.6.** Normalized local auto-correlation function $\Phi_n(1,\tau)$.
$t \in [0,100], \varepsilon = 10^{-7}, \sigma = 5 \cdot 10^5, \delta = 1, p = 1$.

In paper [7] the mechanism of the onset of chaos and its relationship to the characteristics of the spiral attractors are demonstrated for inhomogeneous media that can be modeled by the Ginzburg– Landau equation(4.14). Numerical data are compared with experimental results.

$$\frac{\partial a(x,t)}{\partial t} = i\omega(x)a(x,t) + \frac{1}{2}(1 - |a(x,t)|^2)a(x,t) + g\frac{\partial^2 a(x,t)}{\partial x}, \quad (4.14)$$

$$\frac{\partial a(0,t)}{\partial x} = 0, \frac{\partial a(l,t)}{\partial x} = 0, x \in [0,l], l = 50.$$

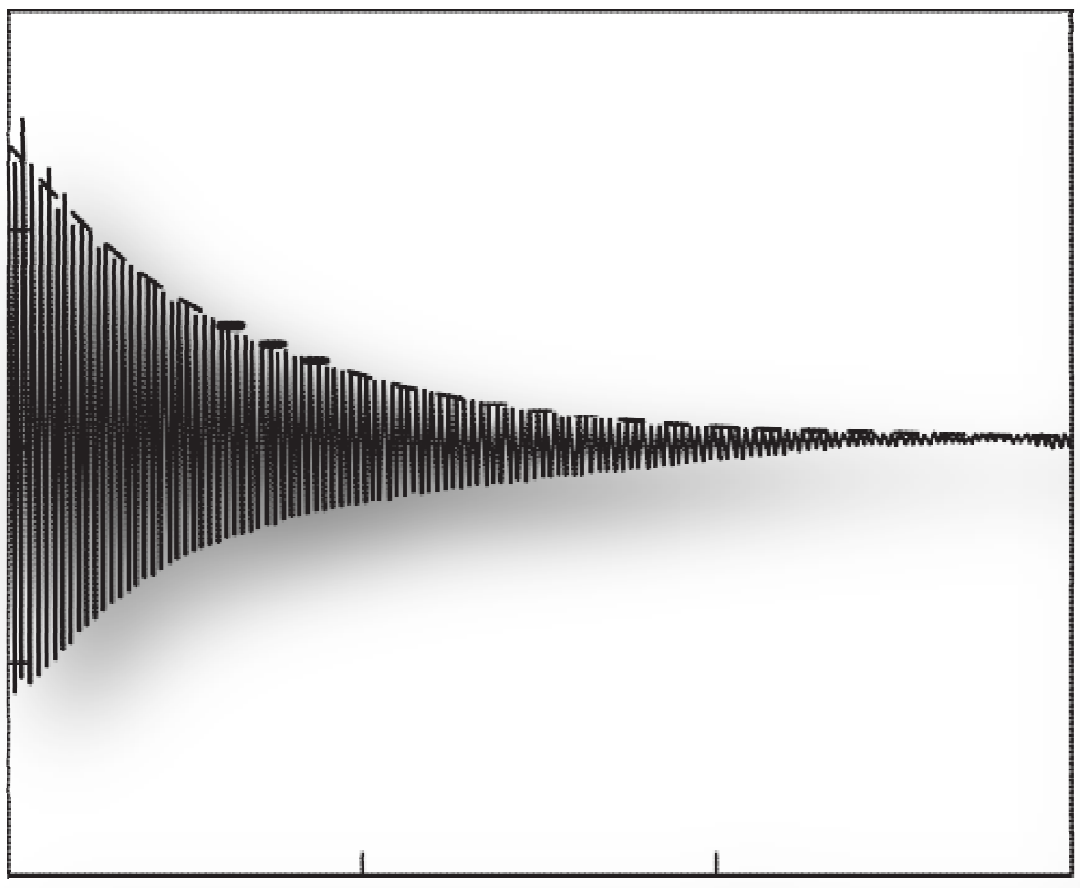

**Pic.4.7.** Normalized local auto-correlation function $\Phi_n(25,\tau)$[7].

However as pointed out above (see Remark1.1-1.4 ) such numerical simulation in fact gives numerical data for stochastic model

$$\frac{\partial a(x,t)}{\partial t} = i\omega(x)a(x,t) + \frac{1}{2}(1 - |a(x,t)|^2)a(x,t) + g\frac{\partial^2 a(x,t)}{\partial x} + \sqrt{\varepsilon}w(x,t), \varepsilon \ll 1, (4.15)$$

$$\frac{\partial a(0,t)}{\partial x} = 0, \frac{\partial a(l,t)}{\partial x} = 0, x \in [0,l], l = 50.$$

## 5.The order of the phase transition from a spatially uniform state to a turbulent state at instant $t \approx 0$.

In order to obtain the character of the phase transition (first-order or second-order on parameters $\varepsilon, p$) from a spatially uniform to a turbulent stateat instant $t \approx 0$one can to use the master equation () of the form

$$\Re(x,t,\varepsilon,\lambda(x,t,\varepsilon)) = 0. \quad (5.1)$$

By differentiation the Eq.(5.1) one obtain

$$\frac{d\Re(x,t,\varepsilon,\lambda(x,t,\varepsilon))}{d\varepsilon} = \frac{\partial\Re(x,t,\varepsilon,\lambda(x,t,\varepsilon))}{\partial\lambda}\frac{d\lambda(x,t,\varepsilon)}{d\varepsilon} + \frac{\partial\Re(x,t,\varepsilon,\lambda(x,t,\varepsilon))}{\partial\varepsilon} = 0. \ (5.2)$$

From Eq.(5.2) one obtain

$$\frac{d\lambda(x,t,\varepsilon)}{d\varepsilon} = -\left(\frac{\partial\Re(x,t,\varepsilon,\lambda(x,t,\varepsilon))}{\partial\varepsilon}\right)\cdot\left(\frac{\partial\Re(x,t,\varepsilon,\lambda(x,t,\varepsilon))}{\partial\lambda}\right)^{-1}.(5.3)$$

Let us consider now 1DEuclidian quantum N-model given by Eq. (4.9)-Eq. (4.10). From corresponding transcendental master equation Eq.(4.13) by differentiation the equation Eq.(4.13) with respect to variable $\lambda$ one obtain

$$\frac{\partial\Re(x,t,\varepsilon,\lambda(x,t,\varepsilon))}{\partial\lambda} = \frac{\{\cos(p\cdot x)-\exp[t\cdot\chi(p)]\cos[p(x-\lambda\cdot\delta\cdot t)]\}\cdot\delta\cdot p}{\chi^2(p)+\lambda^2\cdot\delta^2\cdot p^2} + \frac{\{\cos(p\cdot x)-\exp[t\cdot\chi(p)]\sin[p(x-\lambda\cdot\delta\cdot t)]\}\cdot\lambda\cdot t\cdot\delta^2\cdot p^2}{\chi^2(p)+\lambda^2\cdot\delta^2\cdot p^2} -$$

$$-\frac{\{\cos(p\cdot x)-\exp[t\cdot\chi(p)]\cos[p(x-\lambda\cdot\delta\cdot t)]\}\cdot 2\cdot\lambda\cdot\delta^3\cdot p^3}{[\chi^2(p)+\lambda^2\cdot\delta^2\cdot p^2]^2} - \frac{\{\sin(p\cdot x)-\exp[t\cdot\chi(p)]\cos[p(x-\lambda\cdot\delta\cdot t)]\}\cdot\delta\cdot t\cdot\chi(p)}{\chi^2(p)+\lambda^2\cdot\delta^2\cdot p^2} -$$

$$-\frac{\{\sin(p\cdot x)-\exp[t\cdot\chi(p)]\sin[p(x-\lambda\cdot\delta\cdot t)]\}2\cdot\lambda\cdot\chi(p)\cdot\delta^2\cdot p^2}{[\chi^2(p)+\lambda^2\cdot\delta^2\cdot p^2]^2} + \frac{1}{\sigma}. \qquad (5.4)$$

From Eq.(5.4) for a sufficiently small $t \approx 0$ one obtain

$$\left[\frac{\partial\Re(x,t,\varepsilon,\lambda(x,t,\varepsilon))}{\partial\lambda}\right]_{t\approx 0} = \frac{1}{\sigma}.(5.5)$$

From master equation Eq.(4.13) one obtain by differentiation the equation Eq.(4.13) with respect to variable $\varepsilon$ one obtain

$$\frac{\partial\Re(x,t,\varepsilon,\lambda)}{\partial\varepsilon} = \frac{\left\{-t\cdot\frac{d\chi(p)}{d\varepsilon}\exp[t\cdot\chi(p)]\cos[p(x-\lambda\cdot\delta\cdot t)]\right\}\cdot\lambda\cdot\delta\cdot p}{\chi^2(p)+\lambda^2\cdot\delta^2\cdot p^2} - \frac{\{\cos(p\cdot x)-\exp[t\cdot\chi(p)]\cos[p(x-\lambda\cdot\delta\cdot t)]\}\cdot 2\cdot\frac{d\chi(p)}{d\varepsilon}\cdot\lambda\cdot\delta\cdot p}{[\chi^2(p)+\lambda^2\cdot\delta^2\cdot p^2]^2} +$$

$$+\frac{\left\{-t\cdot\frac{d\chi(p)}{d\varepsilon}\exp[t\cdot\chi(p)]\sin[p(x-\lambda\cdot\delta\cdot t)]\right\}\cdot\chi(p)}{\chi^2(p)+\lambda^2\cdot\delta^2\cdot p^2}+\frac{\{\sin(p\cdot x)-\exp[t\cdot\chi(p)]\sin[p(x-\lambda\cdot\delta\cdot t)]\}\cdot\frac{d\chi(p)}{d\varepsilon}}{\chi^2(p)+\lambda^2\cdot\delta^2\cdot p^2}-$$

$$-\frac{\{\sin(p\cdot x)-\exp[t\cdot\chi(p)]\sin[p(x-\lambda\cdot\delta\cdot t)]\}2\cdot\chi^2(p)\cdot\frac{d\chi(p)}{d\varepsilon}}{[\chi^2(p)+\lambda^2\cdot\delta^2\cdot p^2]^2}(5.6)$$

From Eq.(5.6)for a sufficiently small $t\approx 0$one obtain

$$\left[\frac{\partial\Re(x,t,\varepsilon,\lambda(x,t,\varepsilon))}{\partial\varepsilon}\right]_{t\approx 0}=-\left(t\cdot\frac{d\chi(p)}{d\varepsilon}\right)\frac{\chi(p)\cdot p\cdot\sin(p\cdot x)}{\chi^2(p)+\lambda^2\cdot\delta^2\cdot p^2}=tp^2(p^2-1)^2\frac{\chi(p)\sin(p\cdot x)}{\chi^2(p)+\lambda^2\cdot\delta^2\cdot p^2},(5.7)$$

Therefore from Eq.(5.3),Eq. (5.5) and Eq.(5.7) one obtain

$$[\Im(\varepsilon,x,t)]_{t\approx 0}=\left[\frac{1}{t}\frac{d\lambda(x,t,\varepsilon)}{d\varepsilon}\right]_{t\approx 0}\approx\sigma p^2(p^2-1)^2\frac{\sin(p\cdot x)}{\chi(p)}\mathrm{sign}(p\cdot x)\qquad(5.8)$$

In the limit $t\to 0$ from Eq. (5.8) one obtain

$$\Im(\varepsilon,x)=\lim_{t\to 0}\frac{d\lambda(x,t,\varepsilon)}{td\varepsilon}=\sigma p^2(p^2-1)^2\frac{\sin(p\cdot x)}{\chi(p)}\mathrm{sign}(p\cdot x),\qquad(5.9)$$

and where $\chi(p)=p^2[\varepsilon-(p^2-1)^2]=p^2\rho(\varepsilon),\rho(\varepsilon)=\varepsilon-(p^2-1)^2$.

From Eq. (5.9) follows that

$$\lim_{\rho(\varepsilon)\to 0_+}\Im(\varepsilon,x)=+\infty,\qquad(5.10)$$

$$\lim_{\rho(\varepsilon)\to 0_-}\Im(\varepsilon,x)=-\infty.\qquad(5.11)$$

From Eq. (5.10)-(5.11) follows second order discontinuity of the quantity $\Im(\varepsilon,x,t)$ at instant $t=0$. Therefore the system causing it to make a direct transition from a spatially uniform state $u_{\eta\approx 0}(x,0,\varepsilon)=0$ to a turbulent state in an analogous fashion to the second-order phase transition in quasi-equilibrium systems.

## 6. Chaotic regime generated by periodical multi-modes external perturbation.

Assume now that external periodical force$f(x)$has the following multi-modes form

$$f(x)=-\sum_{k=1}^{m}\sigma_k\sin(p_k x).(6.1)$$

Corresponding transcendental master equation are

$$\Re(x,t,\varepsilon,\lambda)=\sum_{k=1}^{m}\sigma_k\frac{\{\cos(p_k\cdot x)-\exp[t\cdot\chi(p)]\cos[p_k(x-\lambda\cdot\delta\cdot t)]\}\cdot\lambda\cdot\delta\cdot p_k}{\chi^2(p_k)+\lambda^2\cdot\delta^2\cdot p_k^2}+$$

$$+\sum_{k=1}^{m}\sigma_k\frac{\{\sin(p_k\cdot x)-\exp[t\cdot\chi(p_k)]\sin[p_k(x-\lambda\cdot\delta\cdot t)]\}\cdot\chi(p_k)}{\chi^2(p_k)+\lambda^2\cdot\delta^2\cdot p_k^2}+\lambda=0,\chi(p)=p^2[\varepsilon-(p^2-1)^2].(6.2)$$

Let us consider the examples of **QD-**chaotic solutions with a periodical force:

$$f(x)=-\sigma\sum_{k=1}^{m}\sin\left(\frac{kx}{n}\right).(6.3)$$

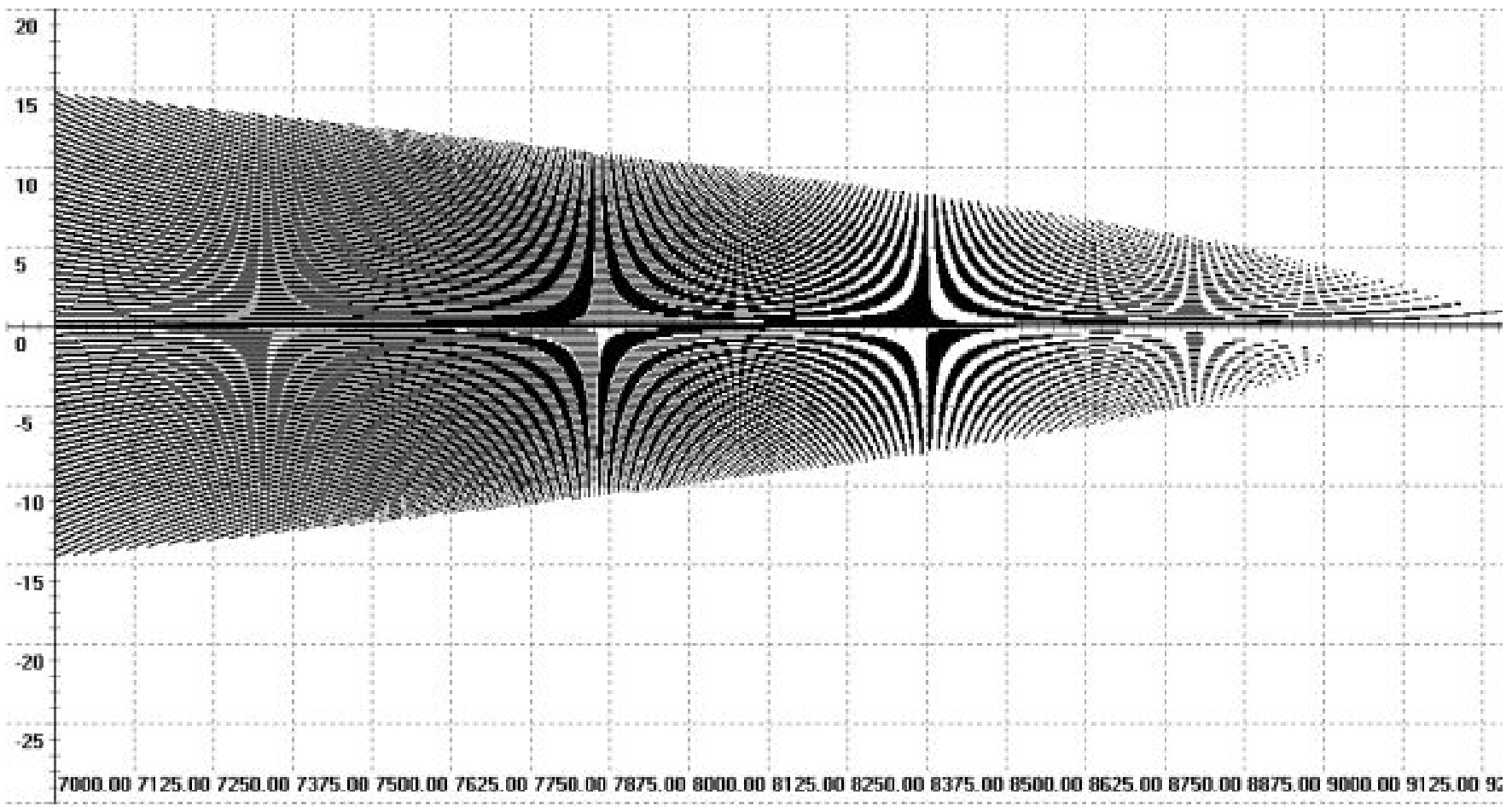

**Pic.6.1.** Evolution of **QD**-chaotic solution $\widetilde{\Re}(10^3,t,\varepsilon)$ in time $t\in[7\cdot10^3,10^4]$, $\Delta t=0.1, m=1, n=100, \varepsilon=-1, p=1, \sigma=10^2, \delta=1, \Delta\lambda=0.01$.

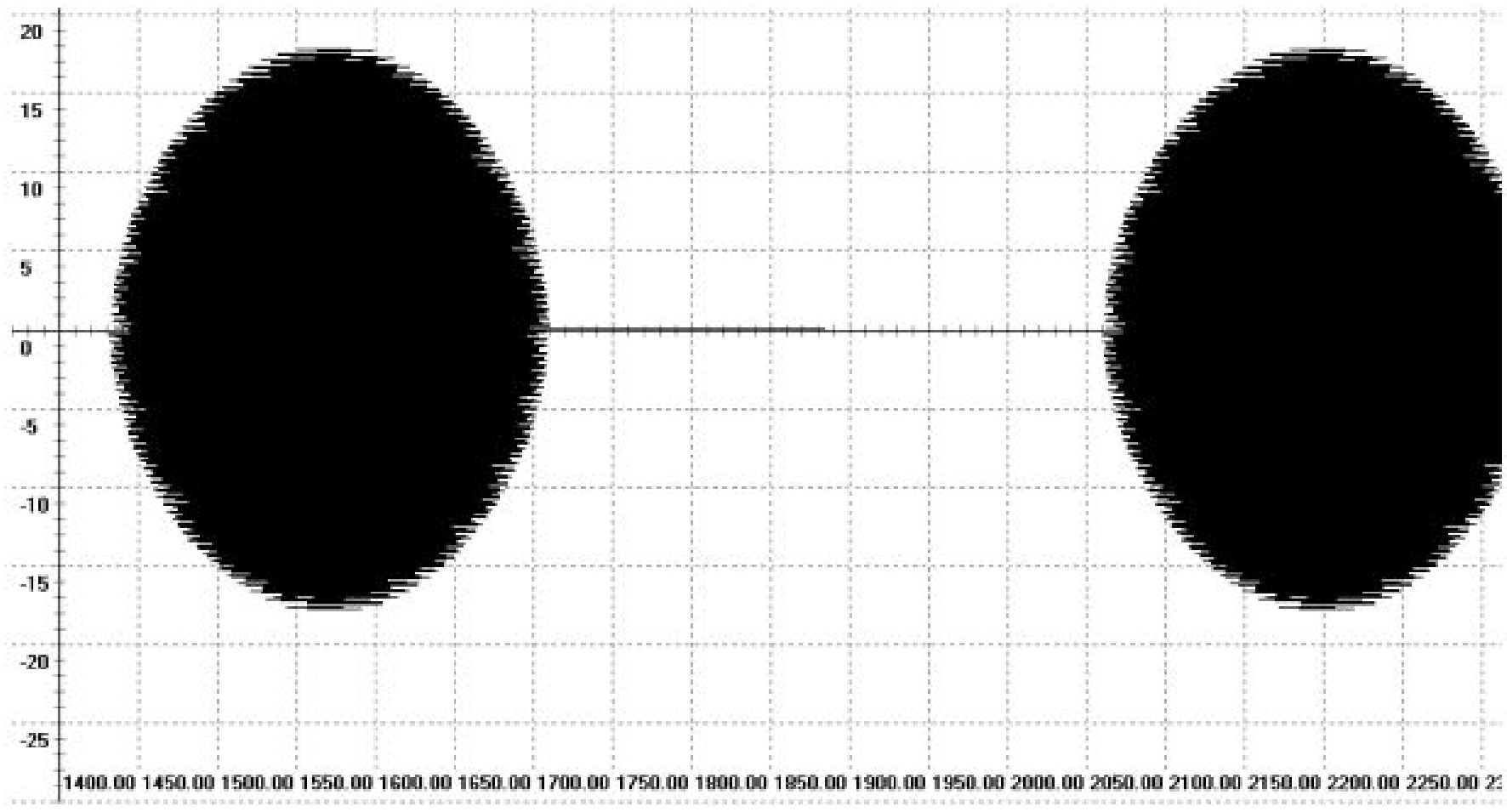

**Pic.6.2.** The spatial structure of **QD**-chaotic solution $\widetilde{\Re}(x,t,\varepsilon)$ at instant $t=10^3$, $x\in[1.4\cdot10^3,2.5\cdot10^3], m=1, n=100, \varepsilon=-1, p=1, \sigma=10^2, \delta=1, \Delta x=0.1, \Delta\lambda=0.01$.

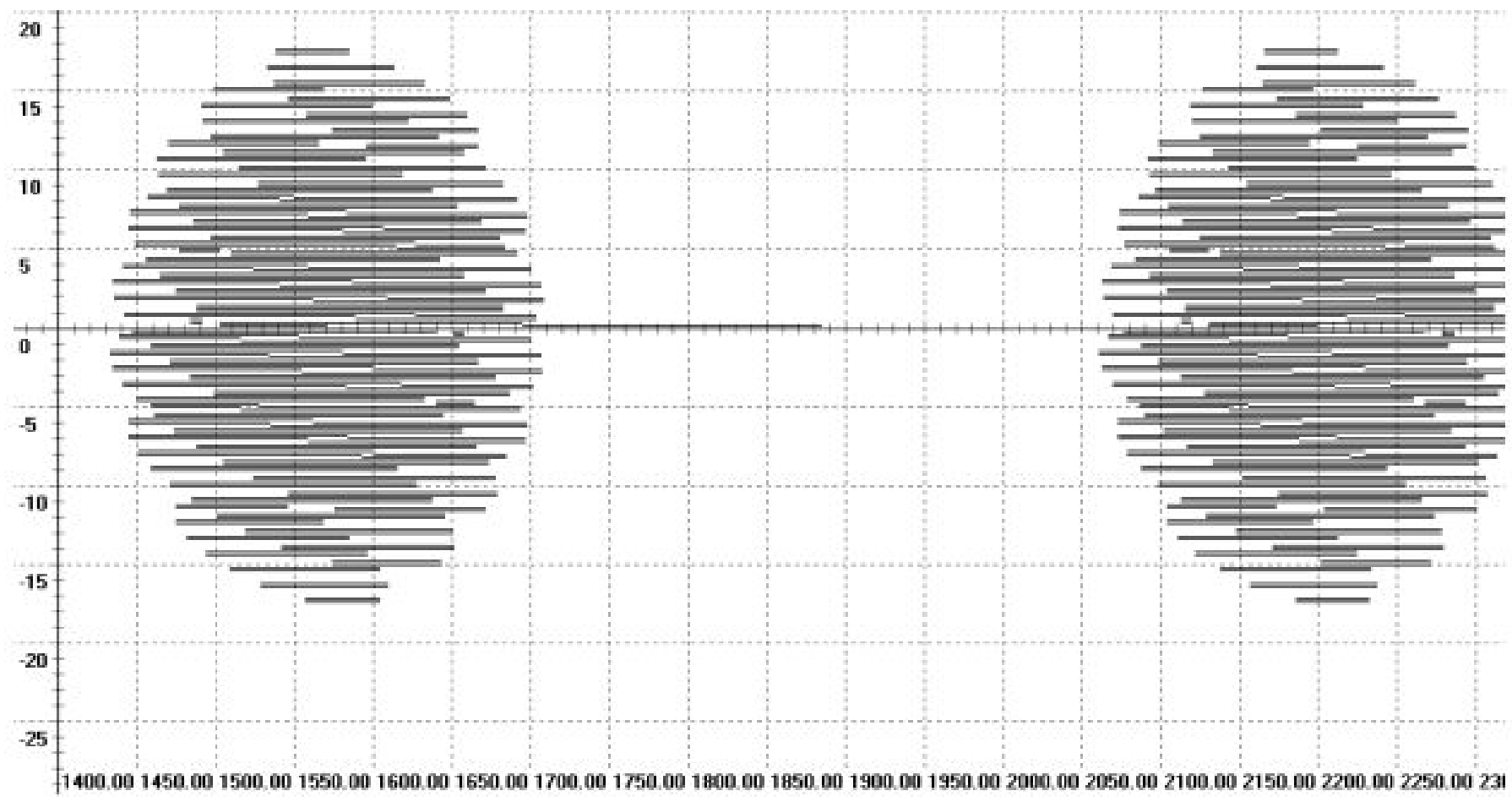

**Pic.6.2.** The spatial structure of **QD**-chaotic solution $\widetilde{\Re}(x,t,\varepsilon)$ at instant $t = 5 \cdot 10^3$, $x \in [1.4 \cdot 10^3, 2.5 \cdot 10^3], m = 1, n = 100, \varepsilon = -1, p = 1, \sigma = 10^2, \delta = 1$, $\Delta x = 0.1, \Delta\lambda = 0.01$.

## 7.Conclusion

A non-perturbative analytical approach to the studying of problem of quantum chaos in dynamical systems with infinite number of degrees of freedom is proposed and developed successfully. It is shown that the additive thermal noise destabilizes dramatically the ground state of the system thus causing it to make a direct transition from a spatially uniform to a turbulent state.

## 8.Acknowledgments

A reviewer provided important clarifications.